\documentclass[aps,prb,twocolumn,superscriptaddress,showpacs,floatfix]{revtex4}

\pdfoutput=1

\usepackage{graphicx}

\usepackage[colorlinks=true,pdfstartview=FitV,linkcolor=blue,citecolor=blue, urlcolor=blue]{hyperref}

\begin{document}

\title{Dynamical thermal response functions for strongly correlated
one-dimensional systems}

\author{Michael R. Peterson} \email{peterson@physics.ucsc.edu}
\affiliation{Physics Department, University of California,  Santa
Cruz, CA  95064, USA}
\author{Subroto Mukerjee} \affiliation{Department of
Physics, University of California, Berkeley, California 94720, USA}
\affiliation{Materials Sciences Division, Lawrence Berkeley National
Laboratory, Berkeley, California 94720, USA}
\author{B. Sriram Shastry} \affiliation{Physics
Department, University of California,  Santa Cruz, CA  95064, USA}
\author{Jan O. Haerter}
\affiliation{Physics Department, University of California,  Santa
Cruz, CA  95064, USA}

\date{\today}

\begin{abstract}
In this article we study the thermal response functions
for two one-dimensional models, namely the Hubbard and spin-less fermion
$t$-$V$ models.  By exactly diagonalizing finite sized systems
we calculate dynamical electrical, thermoelectrical, and thermal
conductivities via the Kubo formalism.  The thermopower (Seebeck
coefficient), Lorenz number, and dimensionless figure of merit
are then constructed which are quantities of great interest to
the physics community both theoretically and experimentally.  We
also geometrically frustrate these systems and destroy integrability
by the inclusion of a second neighbor hop.  These frustrated systems are
shown to have enhanced thermopower and Lorenz number at intermediate
and low temperatures.

\end{abstract}


\maketitle

\section{Introduction}\label{sec-intro}
Strongly correlated electron systems are currently at the
forefront of physics.  These systems provide some of the most
interesting theoretical challenges and  many experimental systems
of fundamental and technological interest consist of strongly
correlated constituent particles\cite{sc_phystoday,sc_rmp}. It is at the
convergence of the fundamental and the technological that thermal
response functions of strongly correlated systems take on
particular importance.  This  is certainly true in the high
thermopower material sodium cobalt
oxide\cite{nco_general,nco_terasaki,ong_nature,cw_prl} as well
as other transition metal oxides such as the high $T_c$
superconductors. Thermal behavior is also of interest in
experimental one-dimensional systems such  as carbon
nanotubes\cite{nanotube}, semiconductor nanowires\cite{nanowire},
and organic compounds\cite{organic-1,organic-2}, to name a few.

One-dimensional systems are very interesting from the point of
view of the exotic collective behavior of excitations due to
reduced dimensionality. This is manifested most strikingly in the
existence of a Luttinger liquid of bosonic excitations in generic
1D systems at low temperatures. Electrical and thermal transport
in these systems is of great interest\cite{kane}. One-dimension also
allows the existence of a class of systems known as integrable
systems, where there is an infinite family of commuting operators
that commute with the Hamiltonian\cite{integ1,integ2,bill-bm}.  
Recently, Shastry~\cite{shastry_1,shastry_2,shastry_3} has
introduced a high frequency formalism for thermoelectrics (discussed 
below) that is particularly suited for strongly correlated electron systems. 
Hence, one-dimensional systems provide good systems in 
which to benchmark his new high frequency formalism.

Geometrical frustration\cite{ramirez} has generated much interest in the
physics community throughout past decades.
Recently it has been shown that strong electron correlations  in
conjunction with the geometrical frustration induced by a
two-dimensional triangular lattice  are the keys to understanding
the Curie-Weiss metallic phase in sodium cobalt oxide\cite{cw_prl}.
Geometrical and/or electrical frustration is also the key to the emergence
of kinetic anti-ferromagnetism\cite{k-afm}, and
to a description of quantum spin glasses\cite{ramirez}.

In a one-dimensional system it is possible to induce frustration by
considering kinetic energy hoppings further than nearest neighbors, i.e., second
nearest neighbors. We demonstrate in this paper
that frustration of this sort also has very interesting effects on transport
properties in addition to the equilibrium properties alluded to
above. One such effect is enhancement of the thermopower
compared to the unfrustrated system 
recently predicted by Shastry~\cite{shastry_1,shastry_2,shastry_3,cw_prl}.   
Understanding this enhancement could lead 
to clues towards the creation, in material science
laboratories, of custom made high thermopower materials with far
reaching consequences to the scientific community.

Theoretically, strongly correlated systems are notoriously difficult
due in part to the fact that perturbation theory
is not applicable.  For dynamical  thermal response
functions (using the Kubo~\cite{kubo} formalism), knowledge of the full
eigenspectrum is necessary, therefore, progress is made via exact
diagonalization of finite sized  systems.  The models and systems
we study are the Hubbard and spin-less  fermion $t$-$V$ model on
one-dimensional rings (periodic boundary conditions) of up to $\mathcal{L}=10$ sites
for the Hubbard model  and $\mathcal{L}=16$ sites for the spin-less
fermion $t$-$V$ model.  Newer theoretical methods are
constantly being developed, such as the dynamical
mean field theory\cite{dmft1,dmft2}, finite temperature Lanczos (FTL)
methods\cite{ftl}, cluster perturbation theory\cite{cpt}, etc..  
In fact, the dynamical mean field theory approximation has 
considered some thermoelectric variables previously, although in a 
limited range of model parameter space, c.f. Ref.~\onlinecite{kotliar_1} 
and \onlinecite{kotliar_2}.  It is important, as a first step, 
to approach these systems with a rigorously 
exact method (exact diagonalization) to,  if nothing
else, provide a benchmark for further approximate studies and
methods. Of course, there are obvious shortcomings to exact
diagonalization, i.e., finite  size effects due to the smallness
of the system and the computational price is often quite expensive.
However, we find only the very low temperature  regimes ($T<1|t|$)
troublesome in our studies which, incidentally, are the regimes that
other approximations also have difficulty tackling.  
Interestingly, the high frequency formalism of Shastry, compared 
to dynamical approximations, is significantly less expensive 
computationally and perhaps could lend itself even 
more profitably to new and existing approximate methods.

We primarily use the Hubbard model for our studies and carry out
the most extensive calculations on it. The $t$-$V$ model is mainly
used to supplement the study of the Hubbard model, partly due to
the fact that some of the transport coefficients turn out to be
identically zero in this model as discussed in Sec.~\ref{sec-tV}
and partly because the calculation of
some of the operators are very involved, owing to the interaction
being between neighboring sites
and not on-site. The numerical calculation of the thermopower is
the most tractable and is performed extensively.  The calculations presented 
here are the first detailed calculations (considering 
the full range of model parameter space) in the literature for the 
thermopower, Lorenz number, and 
dimensionless figure of merit for strongly correlated models.

The plan of this paper is as follows: in section~\ref{sec-models} we
introduce the models to be  studied and provide an overview of our
exact diagonalization procedure.  Section~\ref{sec-kubo} reviews the
Kubo formalism for the thermoelectric conductivities needed to
calculate the physical observables of interest.
Section~\ref{sec-s},~\ref{sec-l}, and~\ref{sec-fom} present
results for the thermopower, Lorenz number and thermoelectric
figure of merit (FOM), respectively, for the models we study. We present
our conclusions and a summary in section~\ref{sec-conc}.

\section{Models}\label{sec-models}

The Hubbard model is described by  a Hamiltonian
with a kinetic energy term which allows
electrons to hop between sites $j$ and $j+\eta$ with probability $t(\eta)$
and an on-site electron repulsion potential energy governed
by parameter $U$, i.e.,
\begin{eqnarray}
\hat{H}&=&-\sum_{j=1}^{\mathcal{L}}\sum_{\eta=1}^{2}
\sum_{\sigma}t(\eta)\{\hat{c}^{\dagger}_{j+\eta\sigma}\hat{c}_{j\sigma}+{\rm h.c.}\}\nonumber\\
&&+U\sum_{j=1}^{\mathcal{L}}(\hat n_{j\uparrow}-1/2)(\hat n_{j\downarrow}-1/2)\;,
\end{eqnarray}
where $\hat{c}^{\dagger}_{j\sigma}(\hat{c}_{j\sigma})$ create(destroys) an electron
with  spin $\sigma=(\uparrow,\downarrow)$ at the lattice
site $j$, $\hat{n}_{j\sigma} = \hat{c}^{\dagger}_{j\sigma}\hat{c}_{j\sigma}$ is the
number operator, and $\eta=1$ and $\eta=2$ for first and second nearest
neighbor hoppings, respectively. The electron operators
obey the usual anti-commutation  rules
$\{\hat{c}_{j\sigma},\hat{c}^{\dagger}_{j'\sigma'}\}=\delta_{jj'}\delta_{\sigma\sigma'}$.
We assume $t(1)=t$, $t(2)=t'$, and $t(\eta)=0$ for all other $\eta$.
With $t'=0$, this Hamiltonian is particle-hole symmetric as can be
seen from the local transformation
$\hat{d}_{j\sigma}^{\dagger}=(-1)^j\hat{c}_{j\sigma}$. This transformation
leaves $t$ and $U$ invariant, while taking $N_\sigma \rightarrow
1-N_\sigma$, where $N_\sigma$ is the total number of electrons
with spin $\sigma$ and the total number
of electrons is $N = \sum_{\sigma}N_\sigma$. Since it takes $t' \rightarrow -t'$, a
non-zero $t'$ breaks particle-hole symmetry. Even in the presence
of $t' \neq 0$, this transformation is useful since it tells us
that a quantity $A(N_\uparrow, N_\downarrow, t, t', U) = \pm
A(\mathcal{L}-N_\uparrow, \mathcal{L}-N_\downarrow, t, -t', U)$. Thus knowing the
value of $A$ up to half-filling for $t'$ and $-t'$ lets us
construct the entire dependence on filling in both cases, provided
we know whether $A$ is odd or even under the particle-hole
transformation. We only consider electron repulsion $U > 0$
and assume periodic boundary conditions (one-dimensional rings).

The spin-less fermion $t$-$V$ models is governed by a
Hamiltonian with a kinetic energy term similar to
the Hubbard model (without the spin) but the potential energy describes a nearest
neighbor repulsion governed by a parameter $V$, i.e.,
\begin{eqnarray}
\hat{H}&=& -\sum_{j=1}^{\mathcal{L}}\sum_{\eta=1}^{2}
t(\eta)\{\hat c^{\dagger}_{j+\eta}\hat c_{j} + {\rm h.c.}\}\nonumber\\
&&+ V\sum_{j=1}^{\mathcal{L}}(\hat n_j-1/2)(\hat n_{j+1}-1/2)\;.
\end{eqnarray}

Like the Hubbard model we assume here that $t(1)=t$,
$t(2)=t'$ and $t(\eta)=0$ for all other $\eta$ and $V>0$. This model too
has the same particle-hole symmetry but without the spin. Once
again, $t'$ breaks the particle-hole symmetry and a calculation up
to half-filling (with $t'>0$ and $t'<0$) lets us extrapolate to all values
of filling.

As with any exact diagonalization there exist computational
limitations due to the obvious difficulties in dealing with
large matrices.  In view of these limitations\cite{cl-note} the Hubbard model is computed
for $N=$1,$\ldots$,5 and 5,$\ldots$,8 electrons on $\mathcal{L}=$10
and 8 site systems, respectively.  These particular systems correspond to
electron fillings (densities) of $n=N/\mathcal{L}=$0.1, 0.2,
0.3, 0.4, 0.5, 0.625, 0.75, 0.875, and 1.  The $t$-$V$ model allows a larger
lattice size stemming from the lack of the spin degree of freedom and we present
data for $L=16$ (and $N=1,\ldots,8$).  The large Hilbert spaces can be reduced
by considering systems with a constant $z$-component of spin (for the Hubbard
model). The sector with the
smallest $z$-component of spin has the largest Hilbert subspace
dimension and dominates the physics.  Linear momentum  is a
good quantum number due to the translational invariance of our
systems and we implement this symmetry to further
reduce the Hilbert space to more manageable proportions.  However,
we point out that the bottleneck of our calculations is not
necessarily the Hamiltonian diagonalization but the
finite temperature averages of certain operators and
evaluation of the full Kubo formulas discussed below.

The density of states of our models for $t'=0$ has two van Hove singularities at the
band edges.  If $|t'|>0.25|t|$ another singularity is introduced
which will invariably produce large changes in the
thermoelectric properties.  However, we are interested here in the
more subtle effects arising from geometrical frustration and strong
interactions, hence, by introducing a non-zero $|t'|<0.25|t|$ the
density of states maintains only the original two van Hove singularities.

In order to display effects of electron correlations the
interaction parameter $U(V)$ must be at least a few times larger than the  bandwidth
which is equal to $W=4|t|$ for both models.  We consider
three values of interaction strength for each model corresponding roughly to the
non-interacting case where the interaction parameter is equal to zero ($U(V)=0\ll W$),
the weakly interacting case where the interaction is equal to roughly the
bandwidth ($U(V)\sim W$), and the strongly interacting
case where the interaction is a few times greater than the bandwidth ($U(V)\gg W$).

\section{Dynamical thermal response functions}\label{sec-kubo}

Conductivities are computed via the Kubo linear response
formalism\cite{mahan} recently presented in
detail by Peterson et al in Ref.~\onlinecite{peterson} which
closely followed the work of Shastry\cite{shastry_1,shastry_2,shastry_3}.
In particular, we calculate the Kubo formulas for the electrical
$\sigma(\omega)$, thermoelectrical $\gamma(\omega)$, and the
thermal $\kappa(\omega)$ conductivities, respectively, which
then allows us to calculate physical quantities of
interest such as the thermopower $S$ (or Seebeck coefficient),  the
Lorenz number $L$, and the dimensionless figure of merit $ZT$, given
as
\begin{eqnarray}
S(\omega,T)=\frac{\gamma(\omega,T)}{\sigma(\omega,T)}\;,
\label{therm}
\end{eqnarray}
\begin{eqnarray}
L(\omega,T)=\frac{\kappa(\omega,T)}{T\sigma(\omega,T)}-S(\omega,T)^2\;,
\label{lorenz}
\end{eqnarray}
and
\begin{eqnarray}
Z(\omega,T)T=\frac{S(\omega,T)^2}{L(\omega,T)}
\label{fom}
\end{eqnarray}
given here for completeness and ease of presentation.  The second term in Eq.~\ref{lorenz} is produced
 by the zero electric
current constraint under a thermal gradient\cite{ziman}.  This term is usually
small (especially at  low temperatures) for metals and semiconductors
and often ignored.  However,  as will be shown, for strongly
correlated lattice systems like those studied here,
this term is an important actor especially at high temperatures.

\begin{figure}[t]
\includegraphics[width=8.5cm]{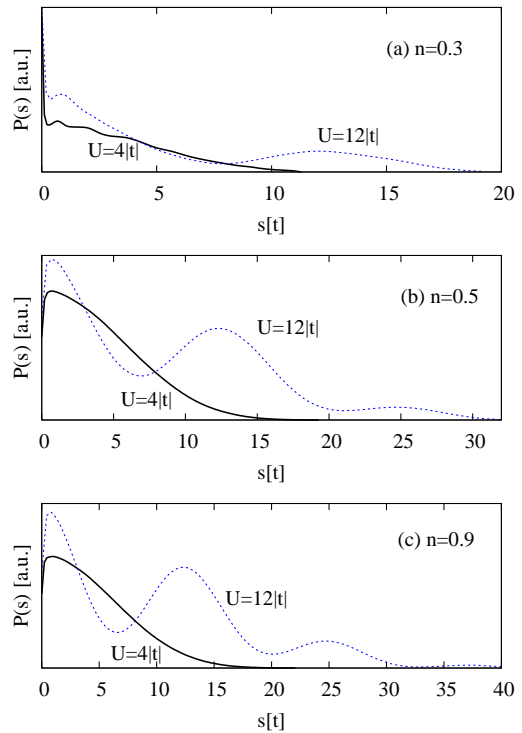}
\caption{Probability density of states $P(s)$ with energy difference
$s=|\varepsilon_i-\varepsilon_j|$.  The  solid black and dashed blue
lines represent $U=4|t|$ and $12|t|$, respectively, for density
$n=0.3$ (a),  $n=0.5$ (b), and $n=0.9$ (c).  The position of the maximum
probability for the lower Hubbard band is insensitive to the
value of $U$.}
\label{eta_plot}
\end{figure}

However, one is usually interested in
the  DC ($\omega\rightarrow 0$) limit of the
dynamical conductivities.  For finite sized systems this requires
the introduction of a level broadening $\epsilon$ which
smoothens the divergences caused by the discrete nature
of the eigenspectrum.  Often $\epsilon$
is taken to be equal to the mean energy level
spacing of the system which is of  order $\mathcal{O}(|t|/\mathcal{L})$.
However, for the Hubbard model, as the interaction energy $U$  is
increased the mean energy level spacing begins to include upper
Hubbard bands.  In Fig.~\ref{eta_plot}(a)-(c) the probability
density of states $P(s)$ with  energy difference
$s=|\varepsilon_i-\varepsilon_j|$ ($\varepsilon_k$ is the
eigenenergy of state $k$) is plotted for three representative cases,
namely, fillings $n=0.3$, $0.5$,  and $0.9$, for $U=4|t|$
(weakly-coupled) and $U=12|t|$ (strongly-coupled).

It is clear from Fig.~\ref{eta_plot}(a)-(c) that the most probable
energy difference between states  in the lower Hubbard band is
relatively immune to changes in either filling or interaction
strength.  However,  the appearance of the upper Hubbard bands in the
strongly-coupled case yields an $\epsilon$ that  is very large
(compared to the bandwidth) and strongly $U$ dependent.  The
large $U$ situation yields a large
$\epsilon$ which tends to mask real physical
contributions to the current matrix elements coming from  transitions
between Hubbard bands.  Therefore, in this work we choose $\epsilon$ to be
approximately equal  to the mean energy level spacing in the lower
Hubbard band for all cases, i.e., $\epsilon=0.8|t|$ for the Hubbard
model\cite{epsilon-foot}.

As discussed previously in detail in Ref.~\onlinecite{peterson}
another frequency limit, besides the DC limit, is the infinite frequency
limit which is defined as (for completeness)
\begin{eqnarray}
S^*(T)=\lim_{\omega\rightarrow\infty}S(\omega,T)=
\frac{\langle \hat{\Phi}_{xx} \rangle}{T\langle \hat{\tau}_{xx}\rangle}\;,
\label{sstar}
\end{eqnarray}
\begin{eqnarray}
L^*(T)=\lim_{\omega\rightarrow\infty}L(\omega,T)=
\frac{\langle\hat{\Theta}_{xx}\rangle}{T^2\langle \hat{\tau}_{xx}\rangle}-[S^*(T)]^2\;,
\label{lstar}
\end{eqnarray}
and
\begin{eqnarray}
Z^*(T)T=\lim_{\omega\rightarrow\infty}Z(\omega,T)T=\frac{[S^*(T)]^2}{L^*(T)}\;.
\label{zstart}
\end{eqnarray}
with the operators $\hat\tau_{xx}$, $\hat{\Phi}_{xx}$, and $\hat{\Theta}_{xx}$
defined in~\cite{shastry_1,shastry_2,shastry_3}.
These quantities can be calculated as equilibrium expectation values of
operators that, while non-trivial, are easier to calculate and
less time consuming numerically than the full dynamical quantities via the Kubo
formulas.  All of the interaction effects remain in these quantities but, importantly,
the dynamics have been separated from the interactions.

\section{Thermopower}\label{sec-s}

The thermopower, or Seebeck coefficient, is defined as the ratio of the thermoelectrical
to the electrical conductivity and measures the
electrical response of a material to a temperature gradient.  Like
the Hall coefficient it is often assumed to measure the sign of the charge
carriers in a particular material.

The thermopower $S(\omega,T)$ can be instructively
rewritten\cite{cw_prl,peterson}, by isolating the
term containing the chemical potential $\mu(T)$, as
\begin{eqnarray}
S(\omega,T)=S_{tr}(\omega,T)+S_{MH}(T)
\label{therm1}
\end{eqnarray}
The first term is due to electrical transport while the second is
entropic in origin and is the familiar Mott-Heikes\cite{chaikin-beni}
term $S_{MH}(T)=-(\mu(T)-\mu(0))/q_eT$.  These two terms both contribute to
the thermopower in different and often competing ways.

Strongly correlated systems are difficult to make theoretical progress on and
the first term in Eq.~\ref{therm1} is usually the most intractable.
For many systems this term is small (especially at low temperatures), compared
to the second term, and fruitfully ignored.  In this approximation the thermopower
is dominated by the Mott-Heikes (MH) term.   The
MH limit is described as the limit when $k_BT \gg |t|$.  This limit is achieved
at either very high temperatures or at more
modest temperatures in narrow band systems\cite{chaikin-beni}.
The usefulness of the MH term is that at high temperatures the chemical potential is
linear in $T$ leaving the MH limit constant.  Even though
the MH limit is constant it contains a non-trivial filing dependence that
arises from the particular nature of the Hilbert space.  Many of
these MH limits were considered previously\cite{chaikin-beni} and below
we quote them for the spin-less fermion $t$-$V$ model for finite and 
infinite $V$ in Eq.~\ref{mh-tv}.  As will be shown below, 
the transport term (first term in Eq.~\ref{therm1}) is
very important for low to intermediate temperatures and the MH term alone is
clearly inadequate in this temperature regime.

The thermopower is expected to vanish in the zero temperature limit.  This
is physically intuitive and has been shown theoretically for the one-dimensional
Hubbard model via a version of the Bethe ansatz solution\cite{stafford}.  In our formalism
this $T=0$ vanishing is accomplished through a subtle balance of the transport
and MH terms as $T\rightarrow 0$.  For thermodynamically large systems this balance is obtained.
However, for finite systems, even
in the non-interacting limit, the balance is not manifest resulting in thermopower divergences
as $T\rightarrow 0$ due to finite size effects.

This balance suggests a rewriting of the thermopower, given
in Ref.~\onlinecite{cw_prl,peterson},
in such a way to ensure that the thermopower vanishes at $T=0$ by forcing
both the transport ($S_{tr}(\omega,T)$) and MH ($S_{MH}(T)$) terms to independently
 vanish in that limit.  This rewriting yields a frequency dependent transport term and
the frequency independent MH term.

Finite sized systems have a few more subtleties which we now describe.  The
chemical potential is commonly defined as
$\mu(T)=\partial F_N(T)/\partial N$,  where $F_N(T)$ is the Helmholtz free energy
for an $N$ particle system.
For our finite systems we will approximate this partial derivative
as $\mu(T)=(F_{N+1}(T)-F_{N-1}(T))/2$ for $N\geq2$
or $\mu(T)=F_{N+1}(T)-F_{N}(T)$ for $N=1$ since there is no system
for $N=0$.  As discussed in Ref.~\onlinecite{peterson} the ground
state degeneracy of a system (if it exists) is discounted when
calculating $\mu(T)$ so as to eliminate a leading order term
linear in $T$ that produces a non-zero $S_{MH}(T)$ as $T\rightarrow 0$.  Further,
finite systems have discrete energy levels giving rise to an energy gap.
The energy gap causes an exponential behavior
in $\mu(T)$ which is not a serious problem since it vanishes faster than $T^2$.  Both
of these particulars, however, create a chemical potential which does not behave
as $T^2$ at low $T$ as expected for thermodynamically large systems.

In the following figures the thermopower will be given in
its ``natural'' units of $k_B/q_e$ where $q_e=-|e|$ with $|e|$ the value
of the electron charge.  For experimental comparison one simply replaces
$k_B/|e|=86\mu V/K$.

\subsection{Hubbard model}

The MH term is the high temperature limit $k_BT\gg |t|$ of $S_{MH}(T)$.  As discussed
elsewhere\cite{chaikin-beni,peterson} the MH limit of the finite $U$ situation is essentially
the uncorrelated band since the temperature is necessarily much larger than $U$.  To
understand the effects strong interactions play (large $U$) one must consider 
infinite $U$ when calculating this limit.

In the finite $U$ case, the MH limit has a single zero crossing at half filling $n=1$ where
the thermopower is seen to change sign and it diverges in the band-insulator
limits ($n=0$ and $n=2$).  For the infinite $U$ case there are two
additional zero crossings and one additional divergence.  The MH limit still diverges
in the band-insulator limits but now also diverges for the Mott insulator ($n=1$).
The two additional sign changes occur for fillings $n=2/3$ and $n=4/3$.
The MH limit in the infinite $U$ case for $n\geq1$ is found from the
MH limit for $0\geq n\geq 1$ through
particle-hole symmetry ($n\rightarrow 2-n$ and $q_e\rightarrow-q_e$).
There is no $t'$ dependence for the MH limits, therefore, they lead to the
conclusion (for $t$ and $t'$) that the Hubbard model should
have interaction induced sign changes of the
thermopower at $n\approx2/3$ (and $n\approx4/3$ using particle-hole symmetry).

\begin{figure}[t]
\begin{center}
\mbox{\bf (a)}
\includegraphics[width=7.cm]{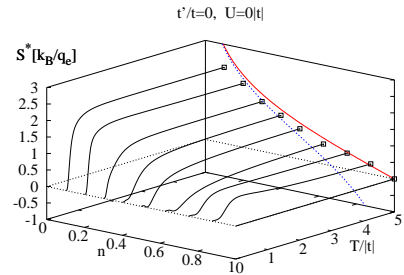}\\
\mbox{\bf (b)}
\includegraphics[width=7.cm]{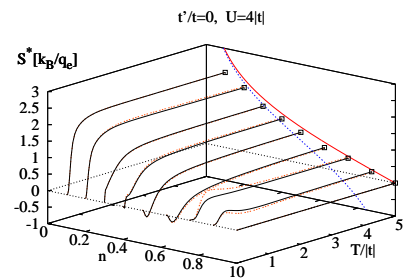}\\
\mbox{\bf (c)}
\includegraphics[width=7.cm]{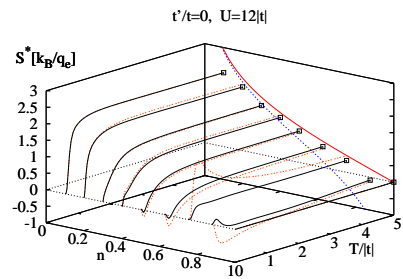}
\end{center}
\caption{(color online) $S^*(T)$ (black line) as a function of filling $n$ and temperature $T$.
For (b) and (c) the orange dotted lines are the
DC limit $S(0,T)$ of the full thermopower for comparison.  Projected
onto the $T=5|t|$ plane are the MH limits for both the finite (red line) and
infinite $U$ (blue dashed line) situations.}
\label{sstar_t0}
\end{figure}

\begin{figure}[t]
\begin{center}
\mbox{\bf (a)}
\includegraphics[width=7.cm]{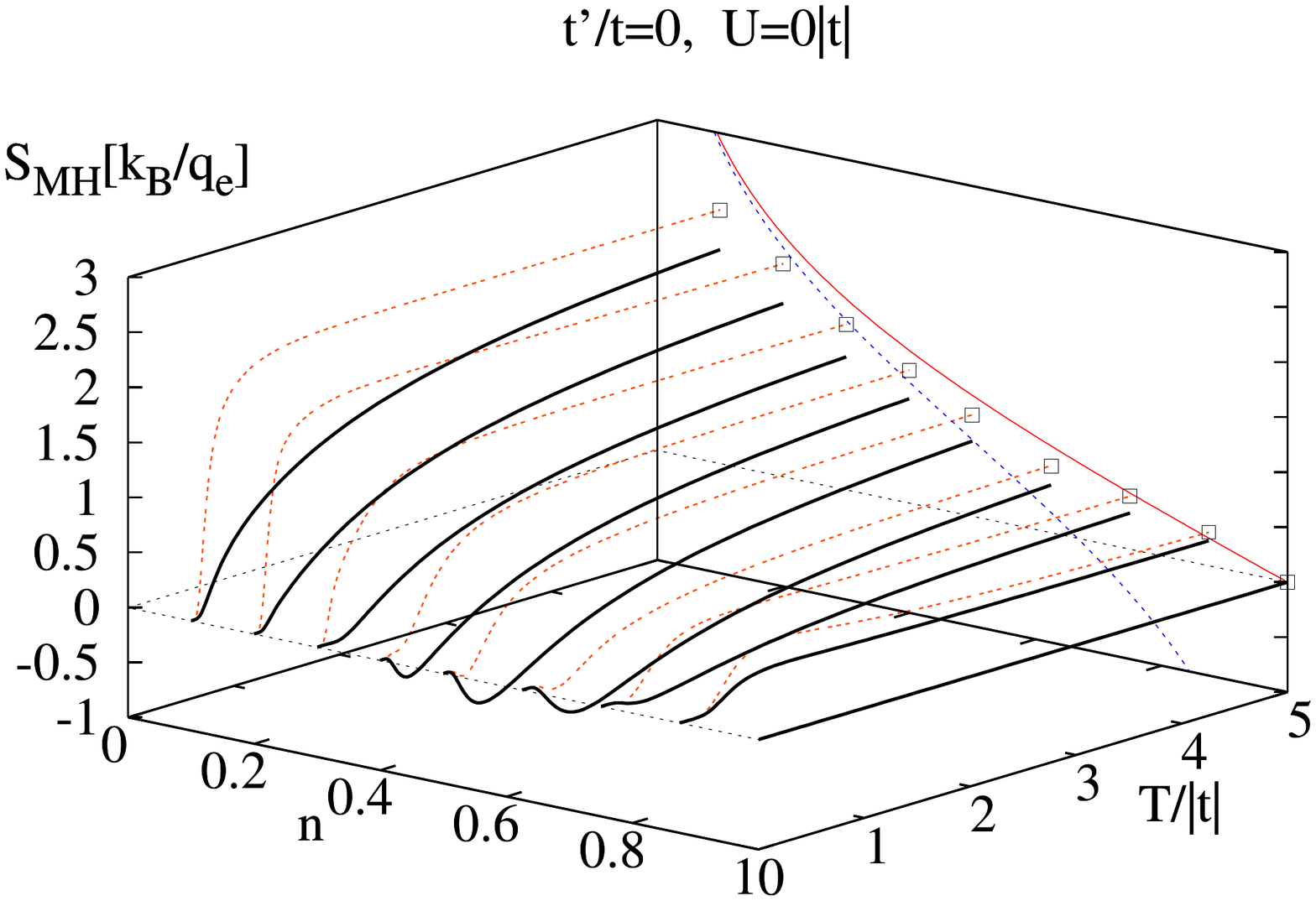}\\
\mbox{\bf (b)}
\includegraphics[width=7.cm]{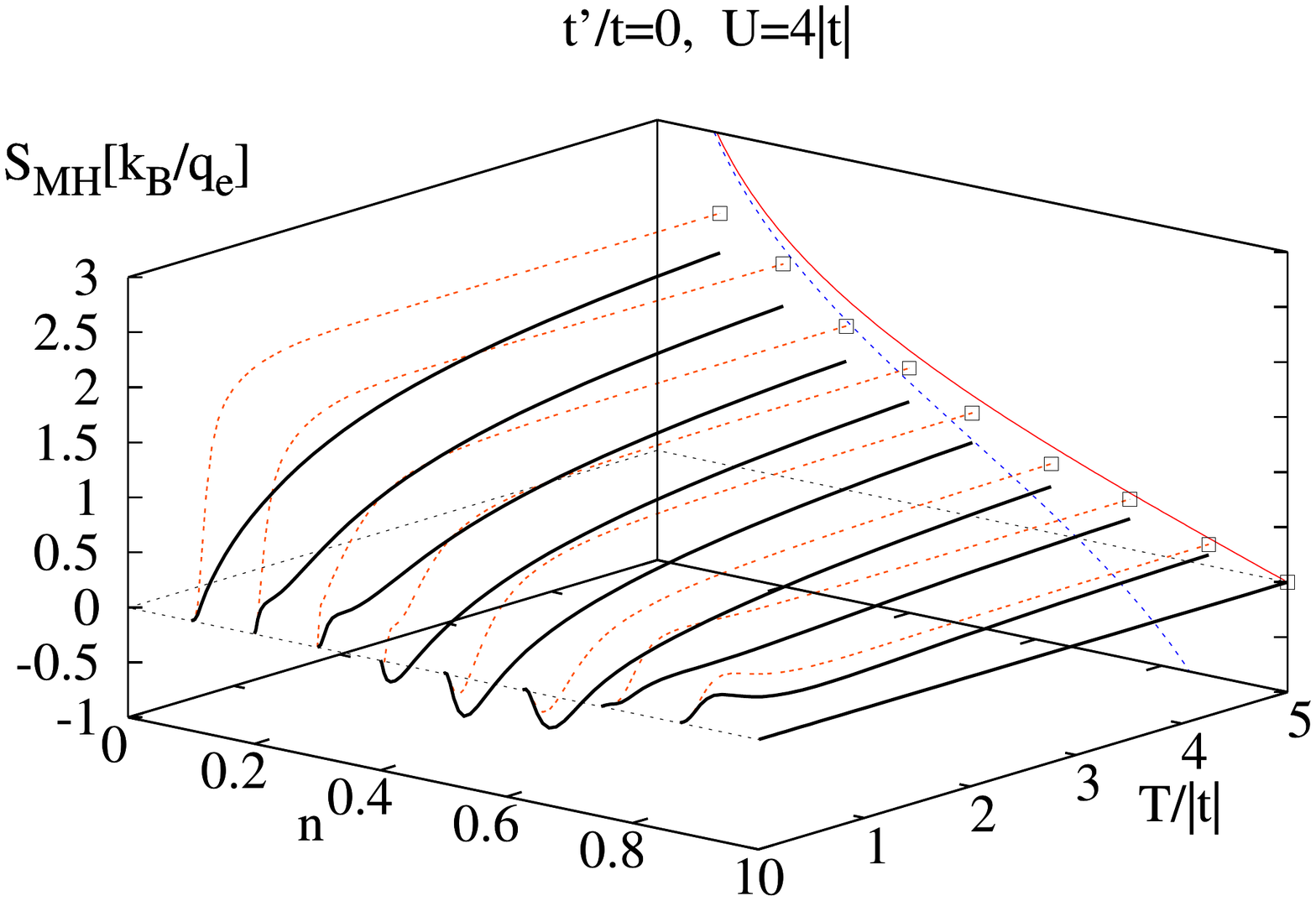}\\
\mbox{\bf (c)}
\includegraphics[width=7.cm]{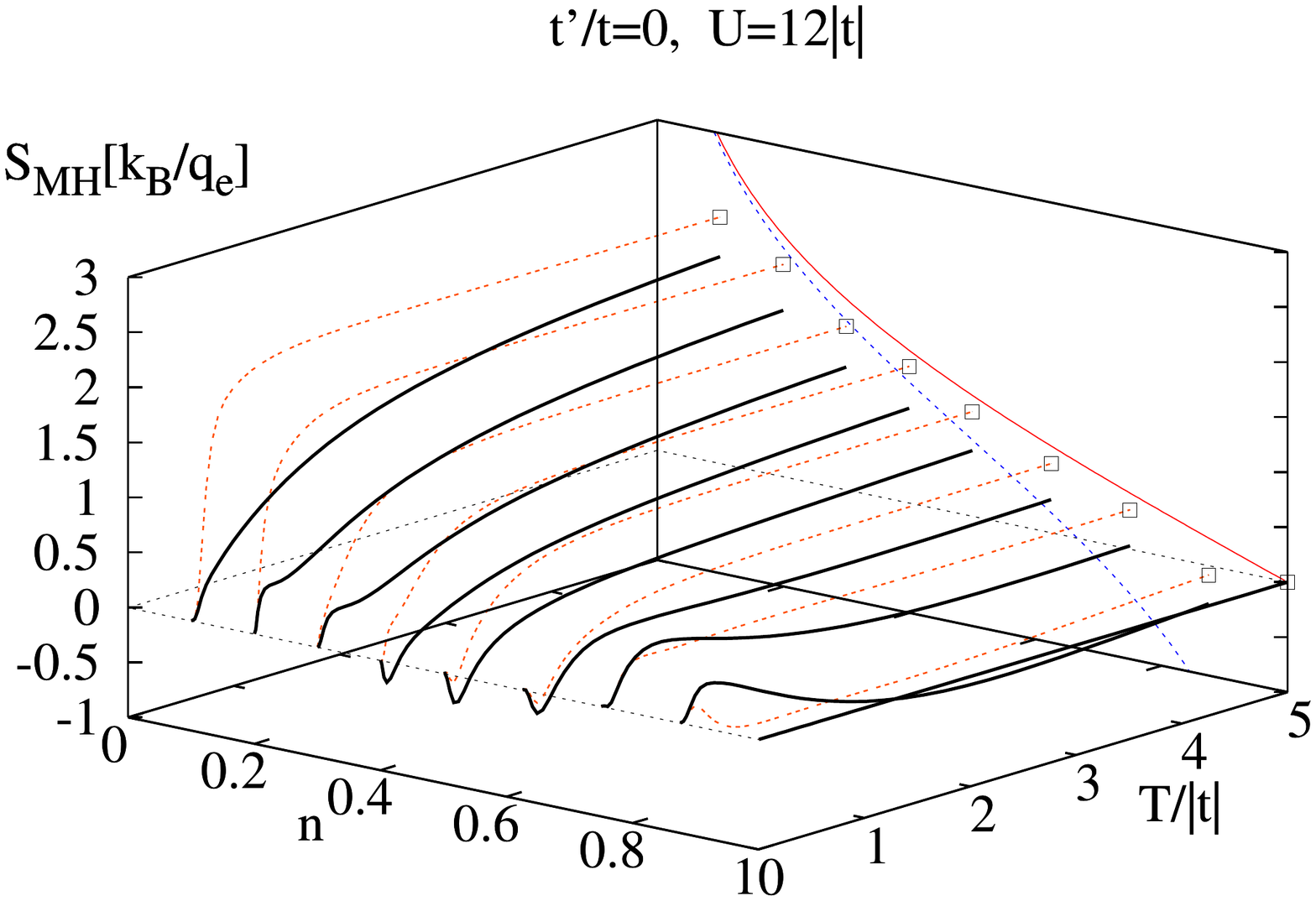}
\end{center}
\caption{(color online) $S_{MH}(T)$ (black line) as a function of filling $n$ and temperature $T$.
For (b) and (c) the orange dotted lines are the
infinite frequency limit $S^*(T)$ for comparison.  Projected
onto the $T=5|t|$ plane are the MH limits for both the finite (red line) and
infinite $U$ (blue dotted line) cases.}
\label{smh_t0}
\end{figure}

First we consider results for the one-dimensional Hubbard model with
$t'=0$.

\subsubsection{Hubbard with $t'= 0$}

Here we consider the Hubbard model with $t'=0$ in the non-interacting ($U=0$), the
weakly correlated ($U=4|t|$), and the strongly correlated
($U=12|t|$) cases, respectively.

Due to the particle-hole symmetry of the $t'=0$ Hubbard
model $\mu(T)$  for $N=\mathcal{L}$ is
\begin{eqnarray}
\mu(T)&=&\frac{1}{2\beta}\ln\left(\frac{\mathcal{Z}_{\mathcal{L}-1}}
{\mathcal{Z}_{\mathcal{L}+1}}\right)\nonumber\\
&=&\frac{1}{2\beta}\ln\left(\frac{\mathcal{Z}_{\mathcal{L}-1}}{e^{-\beta
U}\mathcal{Z}_{\mathcal{L}-1}}\right)
=\frac{U}{2}
\end{eqnarray}
since an energy eigenstate for
 $N$ electrons and $2\mathcal{L}-N$ electrons are related through
$\varepsilon_k(N)=\varepsilon_k(2\mathcal{L}-N)-(\mathcal{L}-N)U$
according to Lieb and Wu\cite{lieb_wu}.
Hence, $S_{MH}(T)$ is identically zero for  all $T$.  The
transport term $S_{tr}(T)$ is also identically zero for all $T$ due to particle-hole symmetry
and hence we find that $S(\omega,T)=0$ for $n=1$ for all $T$ and $\omega$.  This argument
equally applies to the spin-less $t$-$V$ model for $t'=0$.

In Fig.~\ref{sstar_t0}(a)-(c) we plot the infinite frequency limit of the
thermopower $S^*(T)$ and the DC limit ($\omega\rightarrow 0$)
of the full Kubo thermopower $S(\omega,T)$
versus temperature $T$ and filling $n$ for
$U=0|t|$ (a),  4$|t|$ (b), and 12$|t|$ (c).
Projected onto the the $T=5|t|$ plane are the MH limits for finite
and infinite $U$ scenarios.

For the non-interacting case in Fig~\ref{sstar_t0}(a) there is no frequency dependence as
both the charge and heat current operators are diagonal.
The thermopower grows monotonically and,
essentially, linearly from zero at $T=0$ to the MH limit\cite{mh-limit-note} near $T=5|t|$.
As the temperature is increased the slope of the linear
regime is lessened and the thermopower approaches the MH limit more
slowly.  Some finite size effects are evident at the
largest fillings $n=$6/8 and 7/8 calculated where some
somewhat strange low temperature effects are observed.

The effect of interactions shown in Fig.~\ref{sstar_t0}(b)-(c) is to
generally reduce the thermopower for fillings greater than  $n=2/3$.
For fillings below $n\sim 0.625$ the DC and infinite frequency
thermopower are essentially identical.  At fillings greater
than $n=0.625$ $S(0,T)$ and $S^*(T)$ display slight differences at
low temperatures for the weakly correlated regime ($U=4|t|$, Fig.~\ref{sstar_t0}(b))
and marked differences for the strongly interacting regime ($U=12|t|$, Fig.~\ref{sstar_t0}(c)).

The thermopower at half filling in all cases is pinned at zero as discussed.
For $n=0.875$ the effect of interactions is to reduce the
thermopower (Fig~\ref{sstar_t0}(b)), changing its sign (Fig.~\ref{sstar_t0}(c)),
for low to intermediate temperatures.  Eventually
the entropy dominates and the thermopower begins to climb towards its
entropy determined MH limit.
It should be noted that even for the highest temperature $T=5|t|$ the
thermopower remains negative for $n=0.875$ and $U=12|t|$.

In Fig.~\ref{smh_t0}(a)-(c) we plot the MH term of the thermopower ($S_{MH}(T)$)
versus temperature and filling for $U=0$, $4|t|$, and $12|t|$, respectively.
Also shown is $S^*(T)$ for comparison.  Interestingly,
the transport term increases the low to mid temperature
thermopower for all fillings in the non-interacting and
weakly interacting cases.  For the strongly interacting case in
Fig.~\ref{smh_t0}(c) the transport term begins to decrease the
thermopower at low temperatures and high fillings ($n\geq0.75$).

The results in this section can be directly compared with
Ref.~\onlinecite{prelov} which used the  FTL
method which allowed rings of up
to $\mathcal{L}=14$ sites.   Their results were for a much lower
window of temperatures $0.2|t|<T<2|t|$ and show a greater
thermopower suppression for more modest values of $U$
than in the present work.  The thermopower in the low temperature regime is
very sensitive to any changes in either the $S_{tr}(T)$ or $S_{MH}(T)$.
Considering that Ref.~\onlinecite{prelov} used a larger system but an
approximation it is unclear as to the nature of the discrepancy
between those results and the present study.  However, the
qualitative behavior of both calculations is consistent.

\begin{figure*}[t]
\begin{tabular}{cc}
\mbox{\bf (a)}
\includegraphics[width=6.5cm]{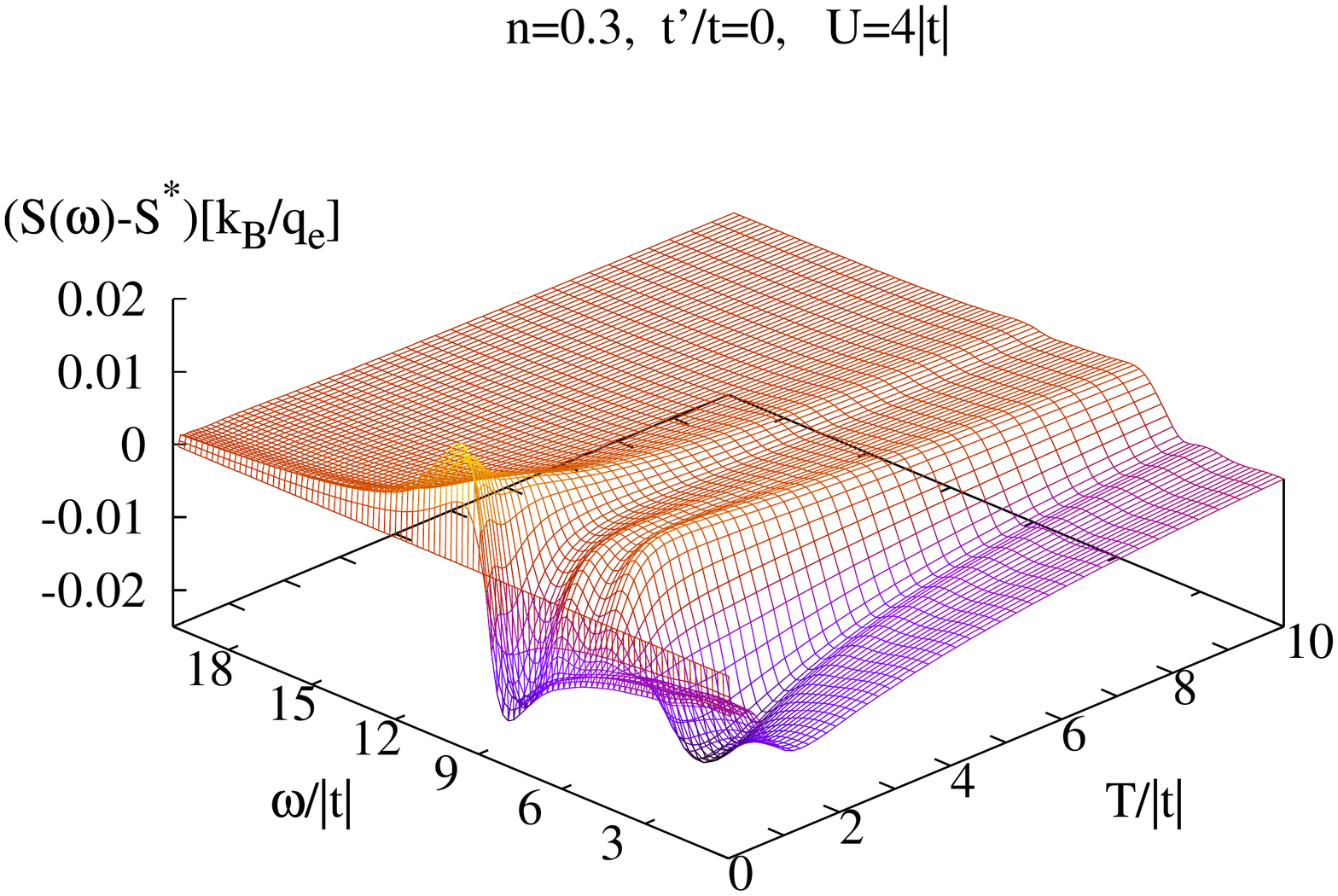}&
\mbox{\bf (b)}
\includegraphics[width=6.5cm]{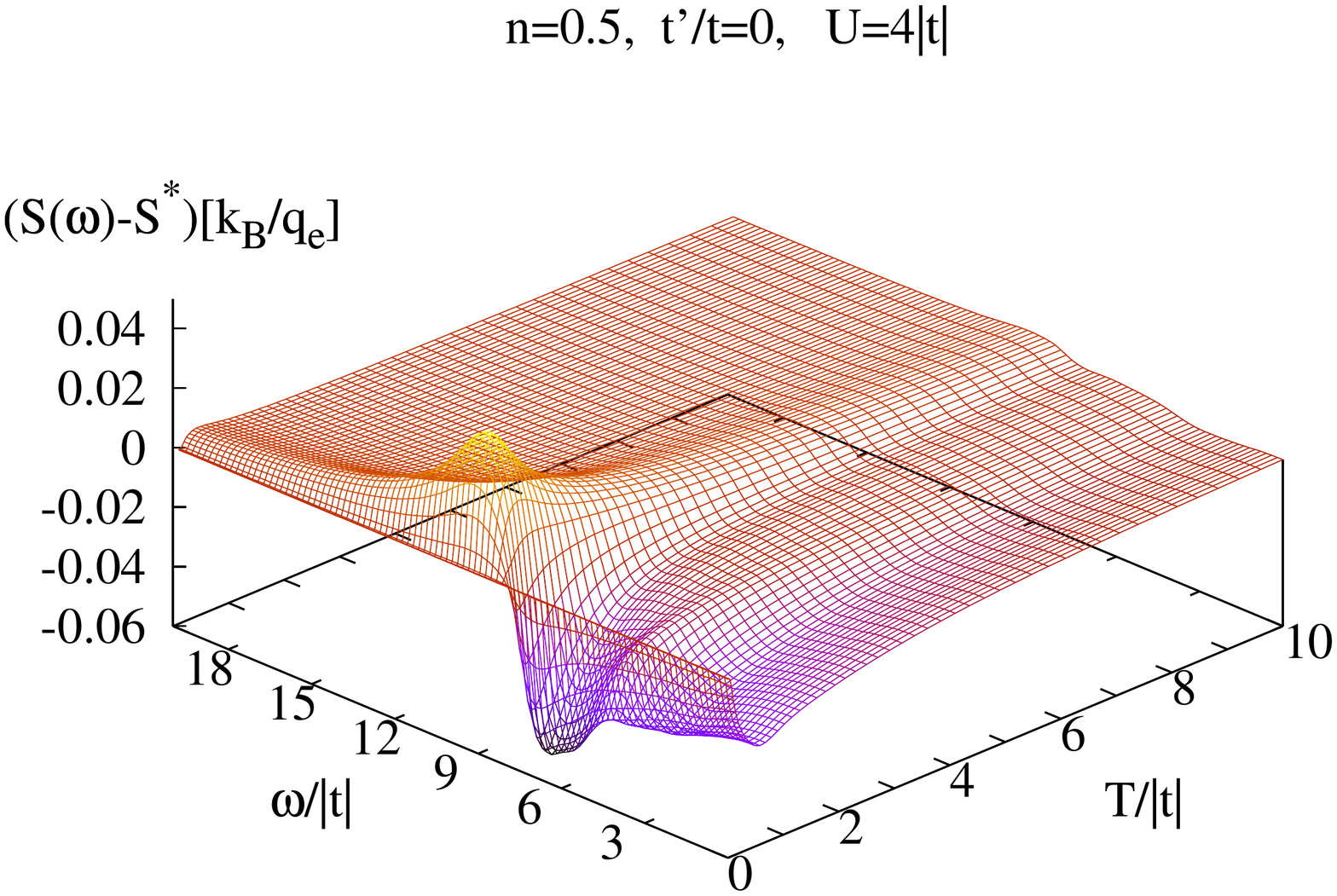}\\
\mbox{\bf (c)}
\includegraphics[width=6.5cm]{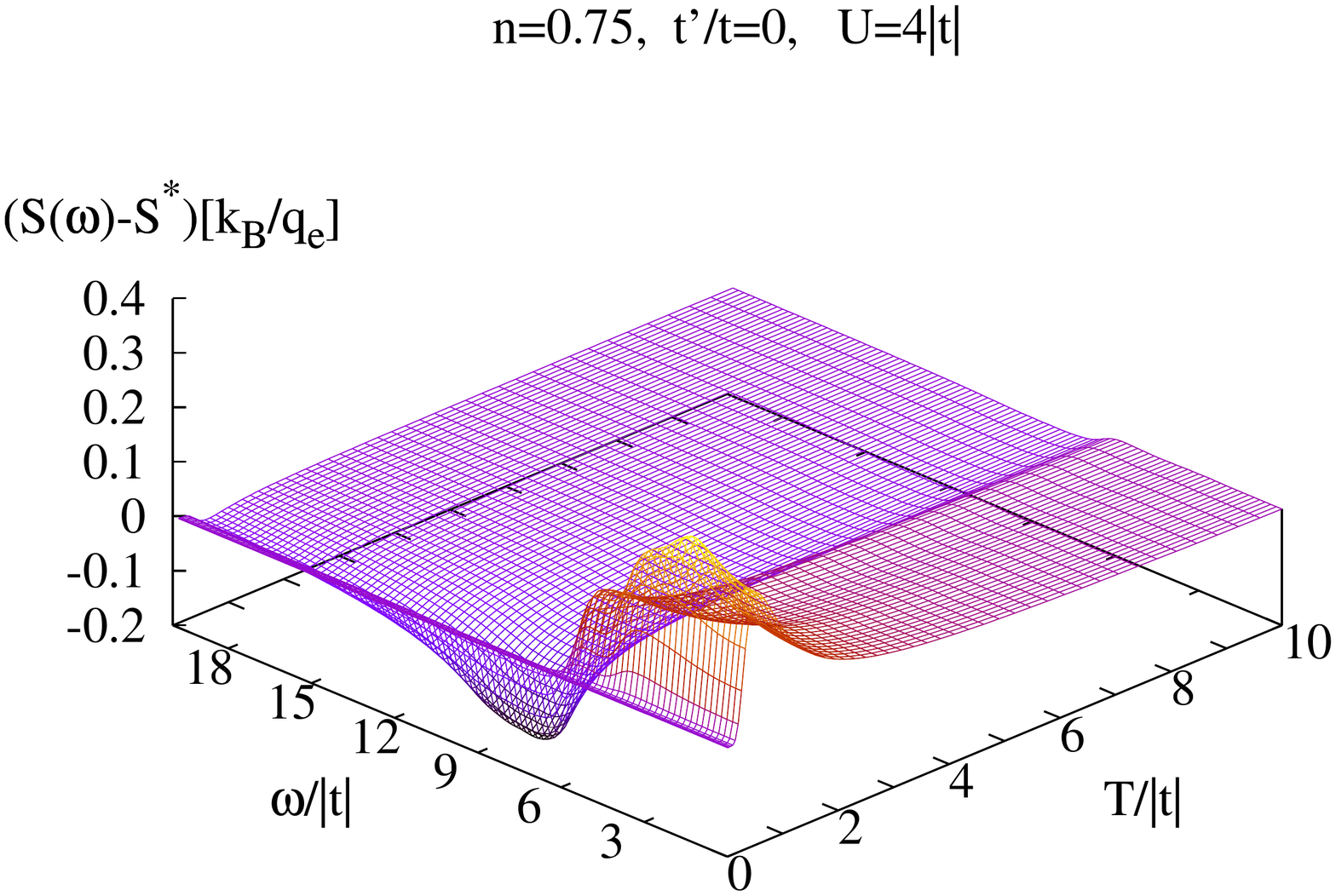}&
\mbox{\bf (d)}
\includegraphics[width=6.5cm]{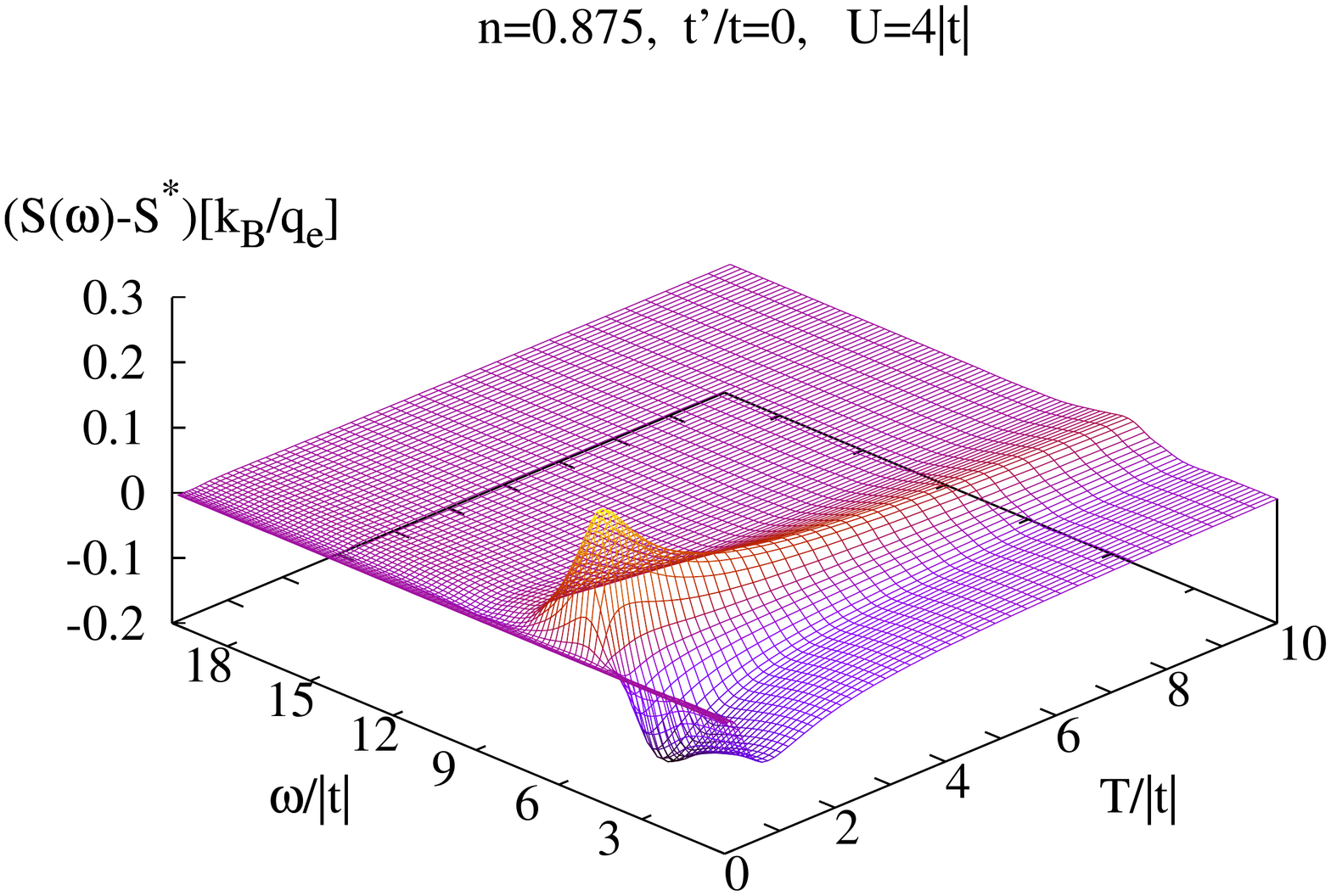}\\
\mbox{\bf (e)}
\includegraphics[width=6.5cm]{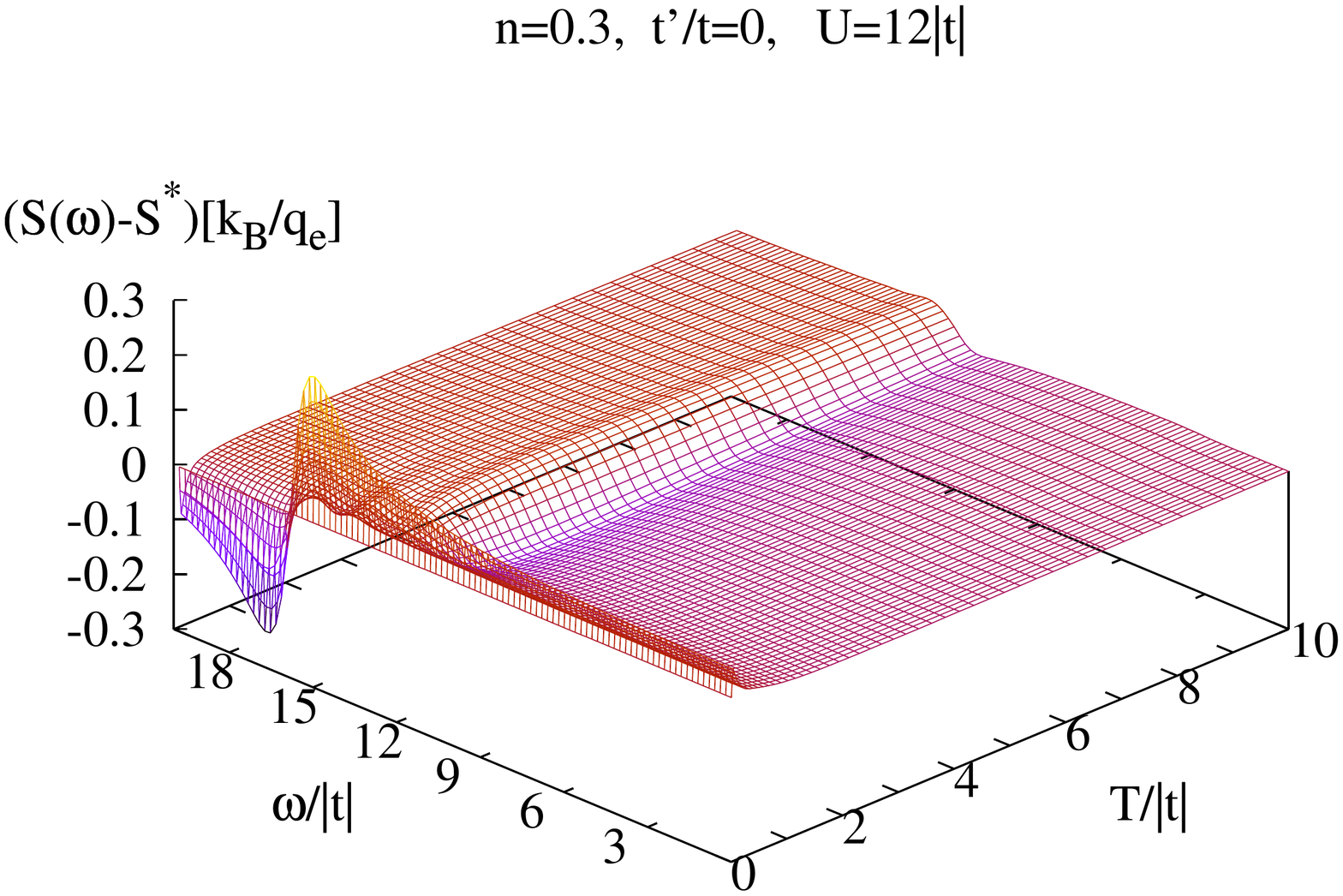}&
\mbox{\bf (f)}
\includegraphics[width=6.5cm]{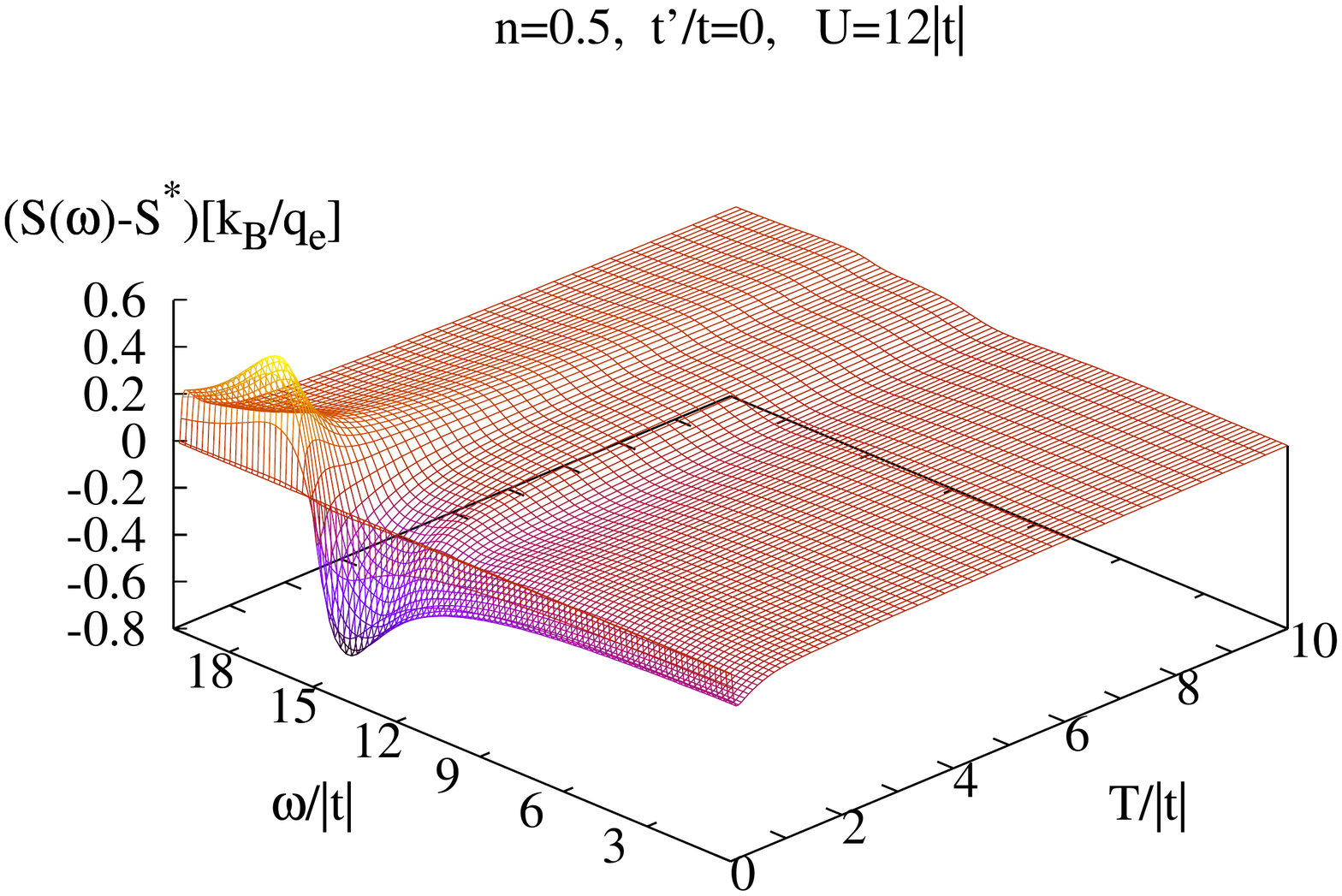}\\
\mbox{\bf (g)}
\includegraphics[width=6.5cm]{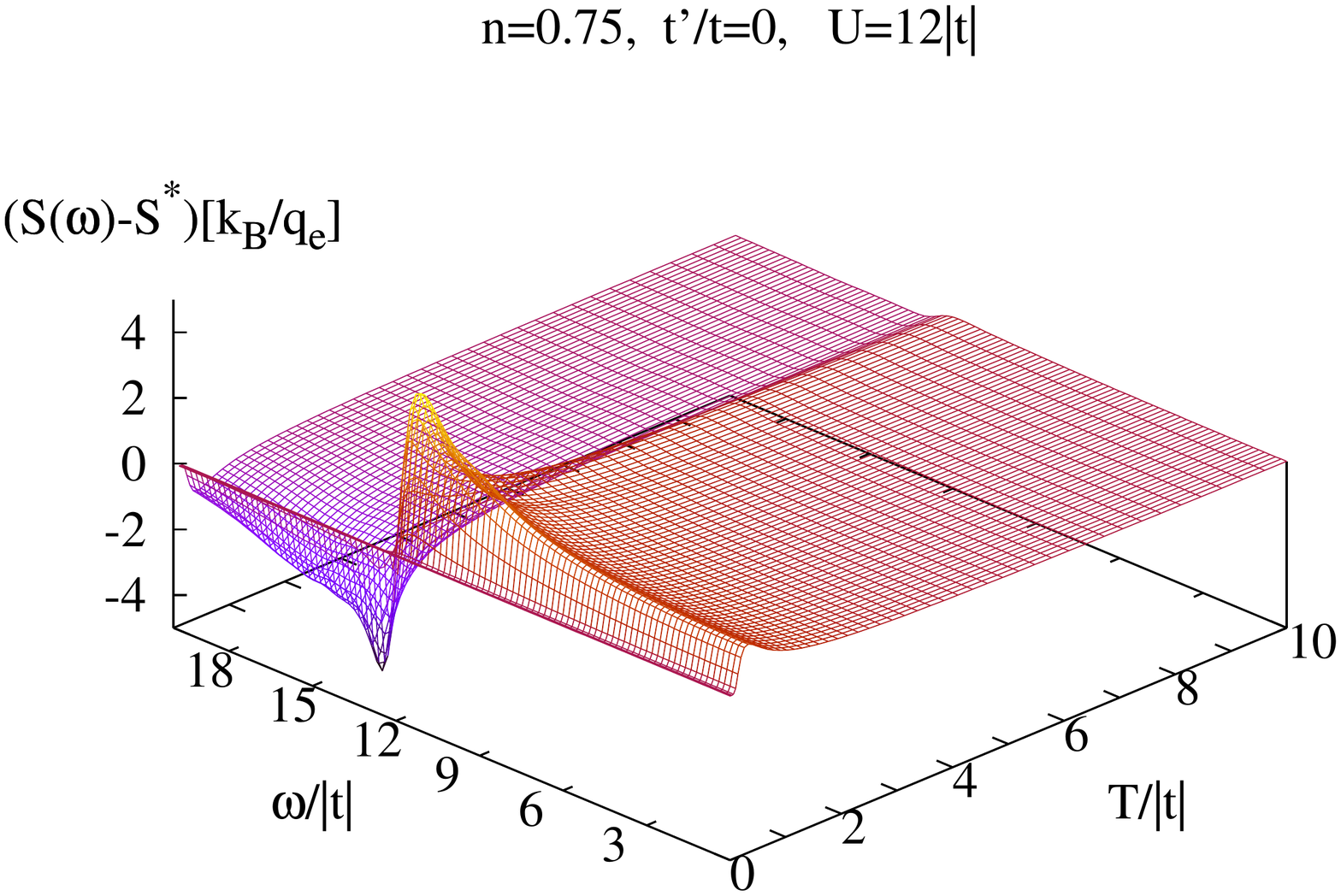}&
\mbox{\bf (h)}
\includegraphics[width=6.5cm]{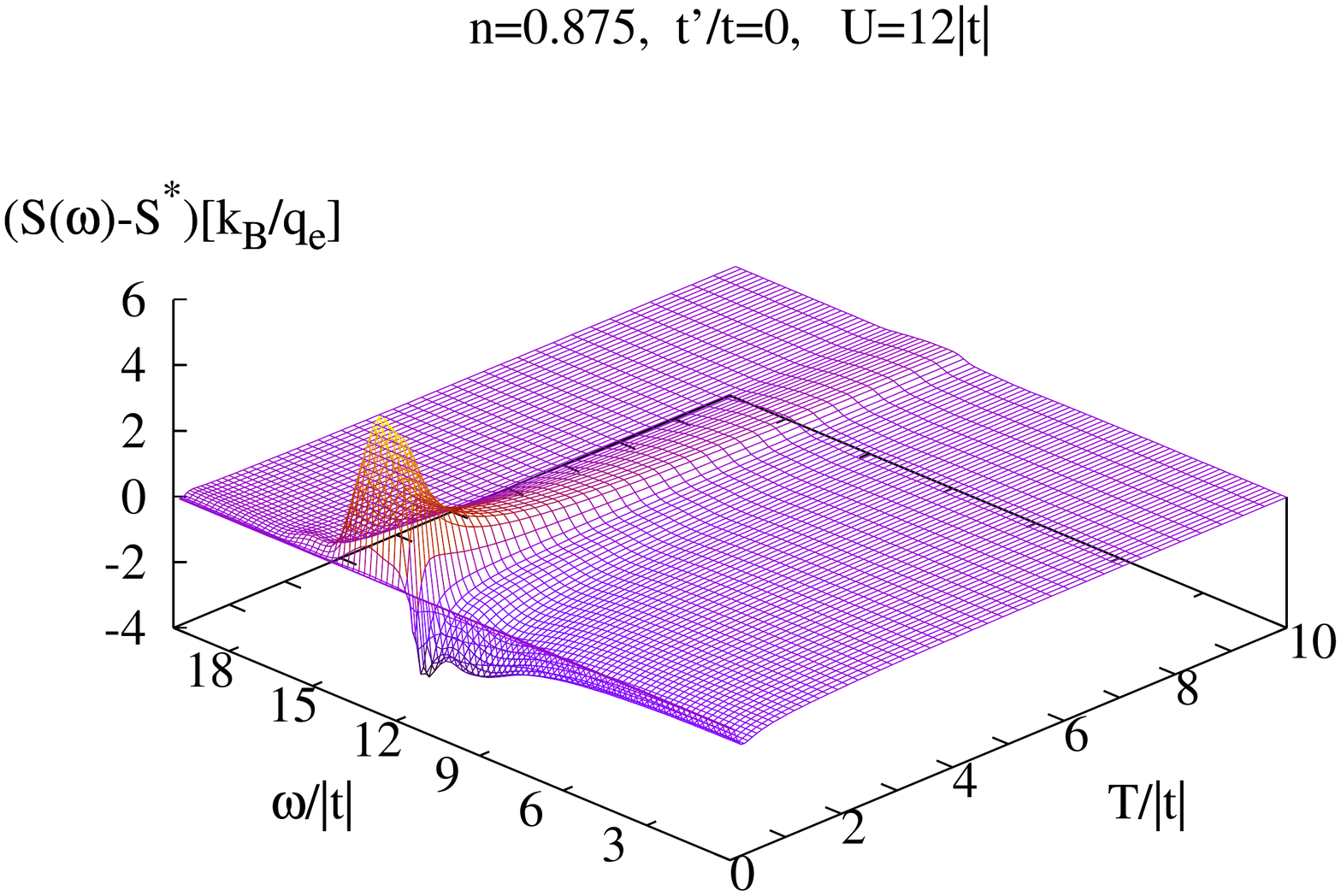}
\end{tabular}
\caption{(color online) $S(\omega,T)-S^*(T)$ as a function of frequency $\omega$ and temperature $T$
for the weak coupling case $U=4|t|$, $t'=0$, and fillings (a) $n=0.3$, (b) $n=0.5$,
(c) $n=0.75$, and (d) $n=0.875$.  Panels (e)-(h) represent
the strong coupling case $U=12|t|$ for the same fillings as the weak coupling case.
The main frequency dependence occurs near $\omega\sim U$ for
low to intermediate temperatures.  The largest frequency dependence near $T=0$ is
likely a finite size effect.  There is no frequency dependence for the
half filled case $n=1$ (not shown).}
\label{s-sstar_t0}
\end{figure*}

Next the full frequency and temperature
dependence of the thermopower is investigated.  Fig.~\ref{s-sstar_t0}(a)-(h)
plots the difference in the full frequency dependent thermopower minus the
infinite frequency limit, i.e.,
$S(\omega,T)-S^*(T)$.  For the weak coupling (Fig.~\ref{s-sstar_t0}(a)-(d))
case there is little frequency dependence except for
when $\omega\sim U$ at low to intermediate temperatures.  In what
is surely a consequence of the finite sized system there appears to be an
even/odd effect in that for even numbers of electrons
(Fig~\ref{s-sstar_t0}(c) and not shown) $S(\omega,T)-S^*(T)$ is positive
for $\omega<U$ and negative for $\omega>U$ while for odd
numbers of electrons (Fig~\ref{s-sstar_t0}(a)-(b)-(d)) the
opposite effect is observed.  Of course, for half filling there
is no frequency dependence (not shown).

\begin{figure*}[t]
\centering
\begin{tabular}{cc}
\mbox{\bf (a)}
\includegraphics[width=7.cm]{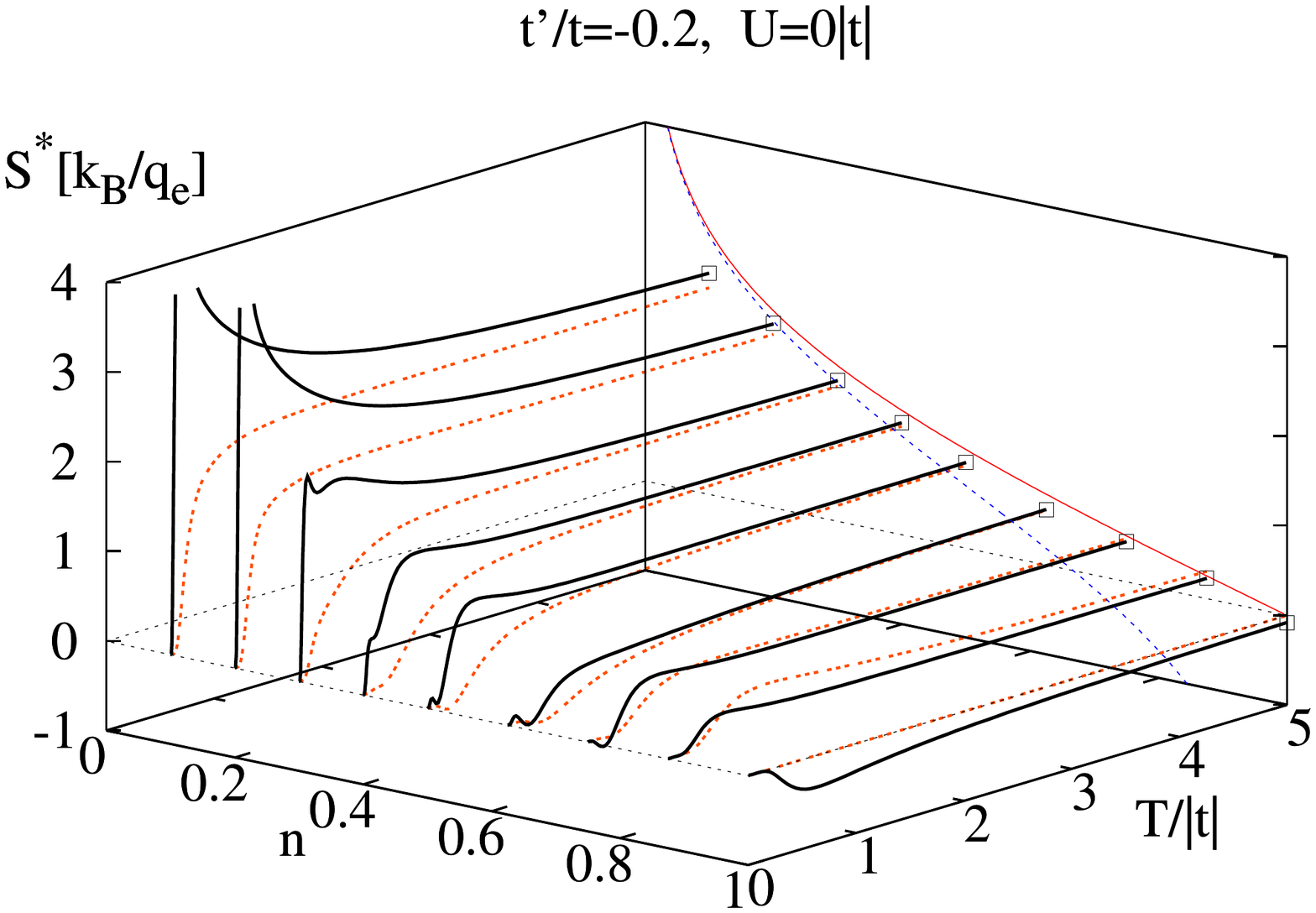}&
\mbox{\bf (b)}
\includegraphics[width=7.cm]{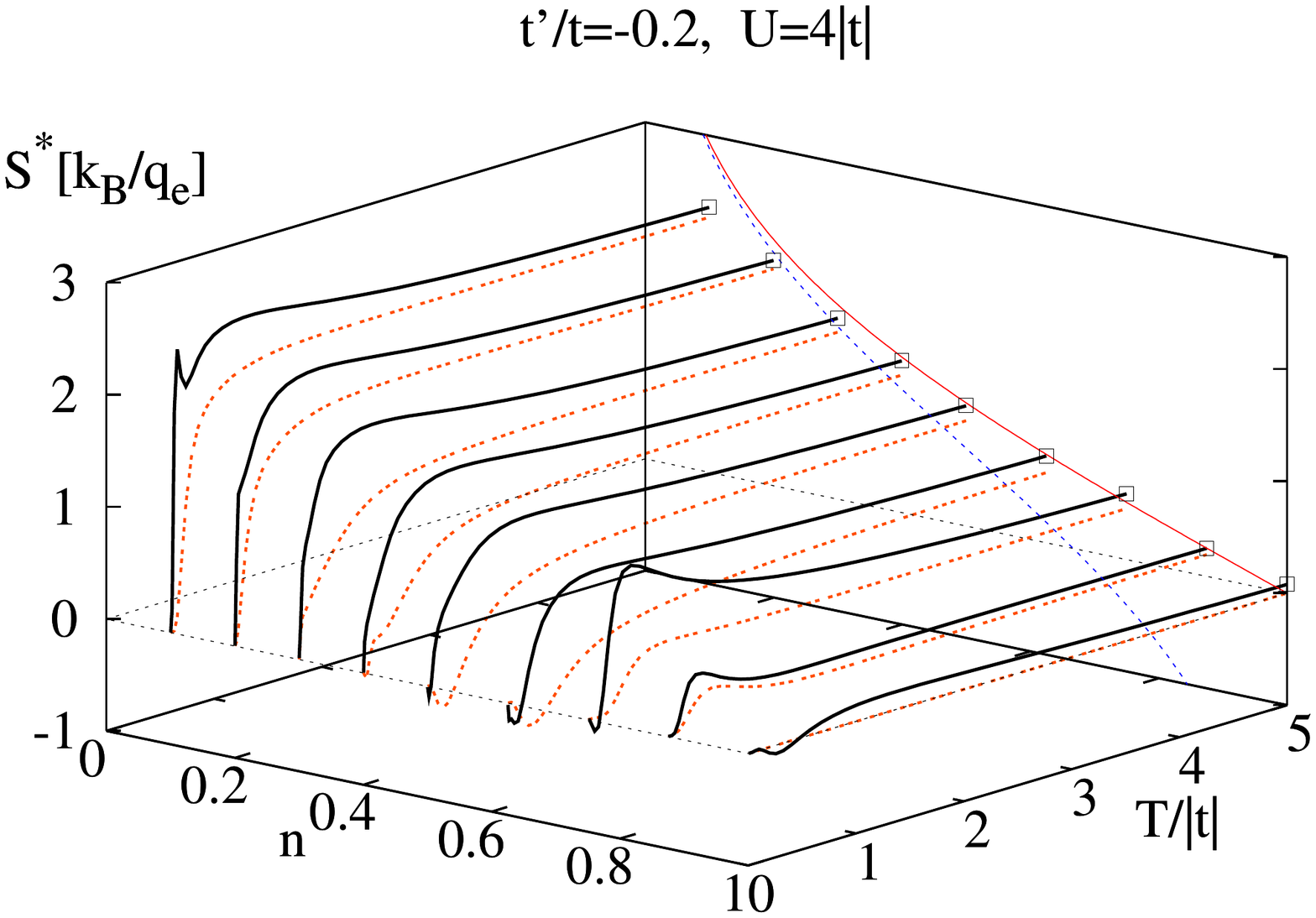}\\
\mbox{\bf (c)}
\includegraphics[width=7.cm]{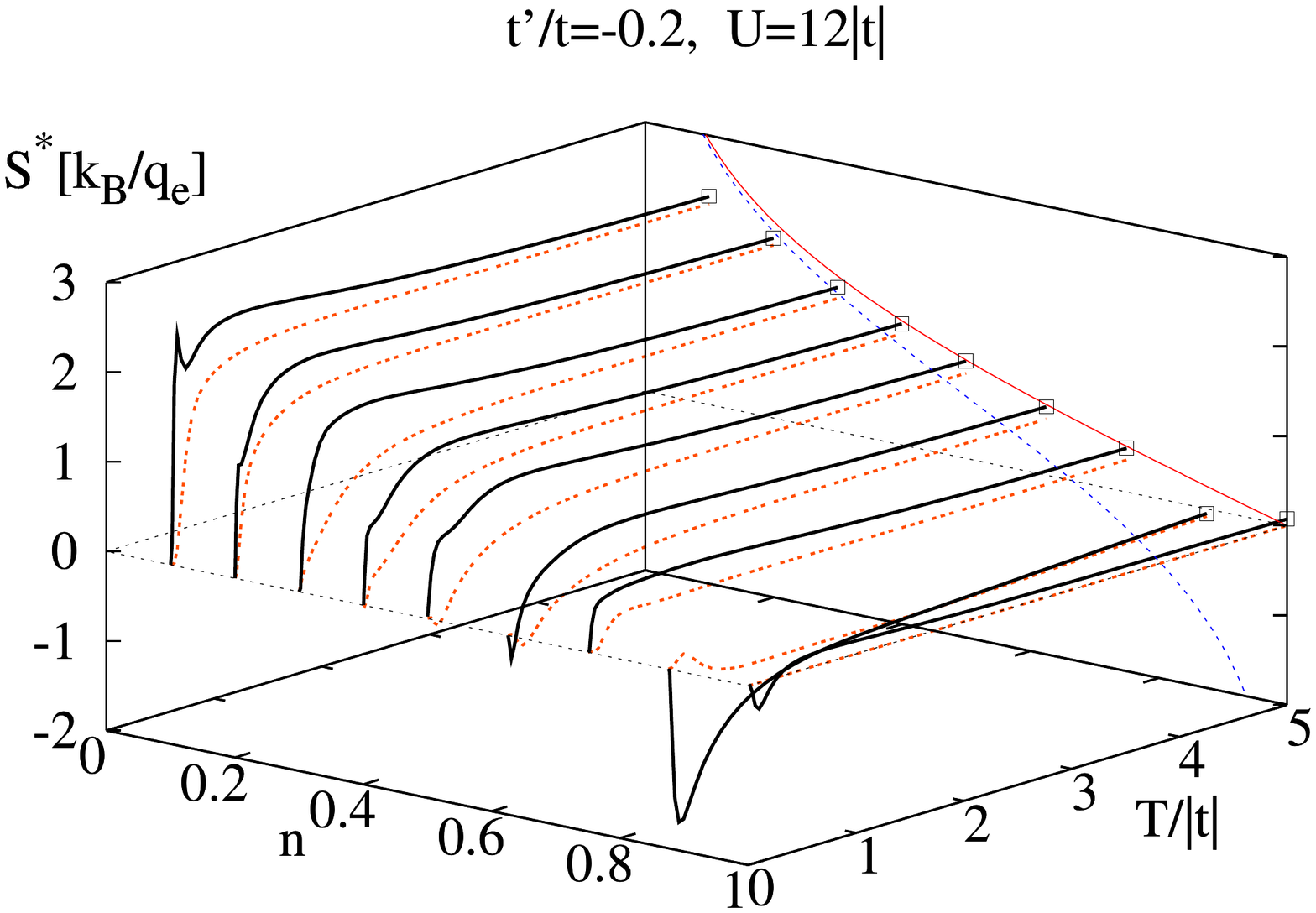}&
\mbox{\bf (d)}
\includegraphics[width=7.cm]{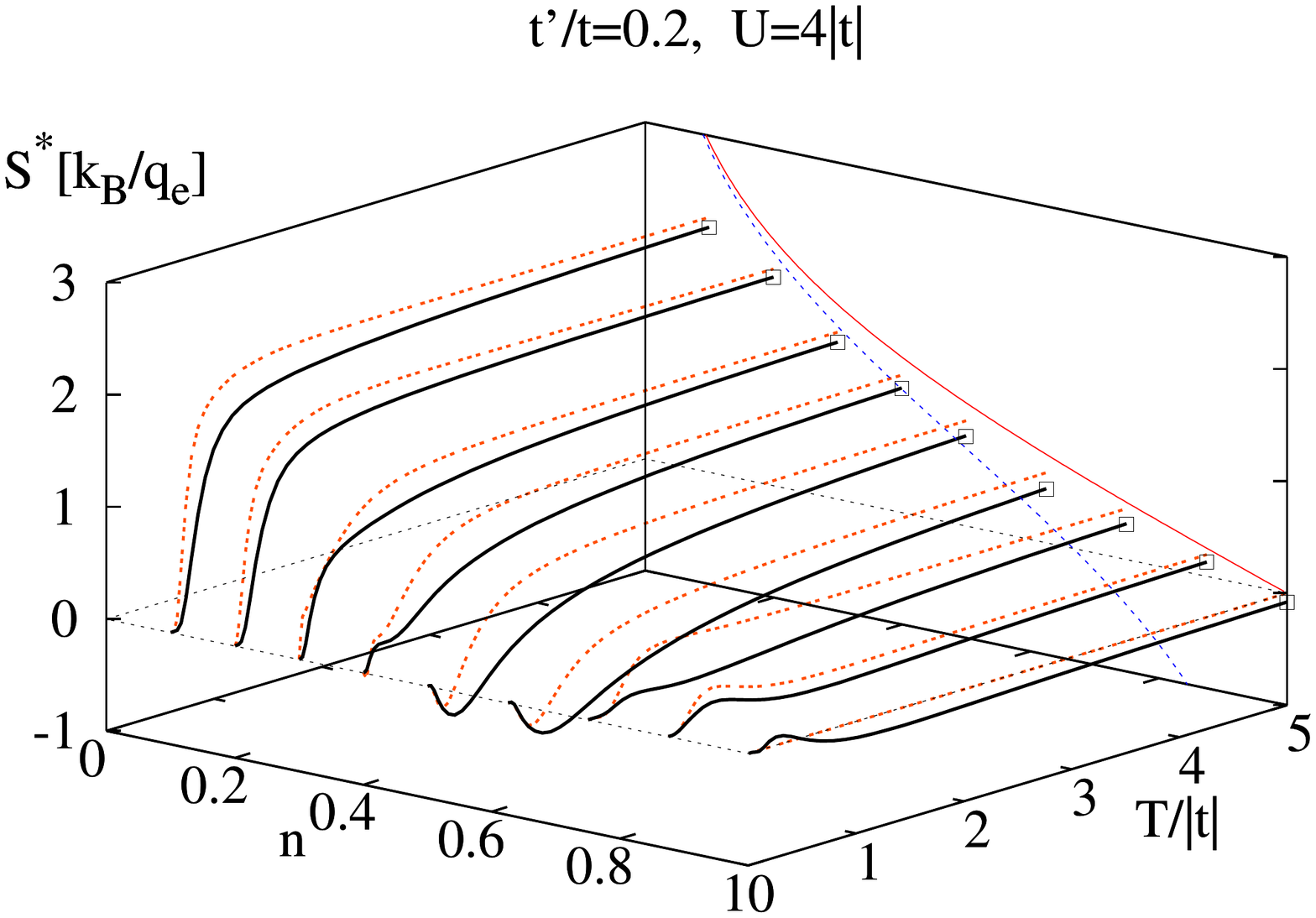}
\end{tabular}
\caption{(color online) $S^*(T)$ (black line) as a function of filling $n$ and temperature $T$ for
(a) $U=0$, (b) $U=4|t|$, and (c) $U=12|t|$ and $t'/t=-02$..  The
full $S(\omega,T)$ (not shown\cite{t02-note}) compares similarly to $S^*(T)$ for $t'\neq 0$ as
for $t'=0$, c.f. Fig.~\ref{s-sstar_t0}.
The red curve is for $t'=0$ to facilitate an easy comparison.  As
an example of the other sign of $t'$, namely, $t'/t=0.2$, (d) shows the expected
reduction in $S^*(T)$ for $U=4|t|$ (similar results obtained for $U=0$ and
$U=12|t|$ are not shown).  The MH limits are projected unto the $T=5|t|$ plane
for finite $U$ (red line) and infinite $U$ (blue dotted line).}
\label{sstar_t-02}
\end{figure*}

The same qualitative behavior is found for the strongly correlated
regime in Fig.~\ref{s-sstar_t0}(e)-(h).  Again the even/odd effect is
obtained (cf. Fig.~\ref{s-sstar_t0}(g) vs. Fig.~\ref{s-sstar_t0}(e), (f), and (h))
and the frequency dependence is generally weak except
for $\omega\sim U$.  The main difference between the $U=12|t|$ and the
$U=4|t|$ cases  is that the frequency dependence at $\omega\sim U$
in the former is larger.  It should be kept in mind that the largest
frequency dependences occur for the lowest temperatures and the thermopower
for a finite sized system in the low temperature regime is susceptible
to finite sized effects that could possibly not survive in the
thermodynamic limit.  Therefore, it is safe to assume that our
results are qualitatively representative of a thermodynamically
large Hubbard model
but perhaps not quantitatively accurate.

Similar qualitative frequency dependence was seen in Ref.~\onlinecite{prelov} for
a low temperature slice ($T=0.5|t|$) and high fillings ($n>0.7$).
The difference in that work and the present work is that the frequency dependence occurred
nearer to $\omega\sim U/2$ in the former where in the present result
the largest frequency dependence occurs near $\omega\gtrsim U$.

Unless one is
concerned with extremely low temperatures and frequencies
similar in magnitude to the interaction strength $U$ the infinite
frequency thermopower $S^*(T)$ is a good representative of the full
thermopower $S(\omega,T)$.  Recently, a similar
calculation on the strongly correlated $t$-$J$ model
for a two-dimensional triangular lattice was carried out
in Ref.~\onlinecite{peterson} where it was found that the frequency
dependence was much weaker justifying the use of $S^*(T)$
in place of $S(\omega,T)$.

\subsubsection{Hubbard with $t'\neq 0$}\label{hub-t-02}

Here we investigate the effect of frustration on the thermopower by considering
the case of a non-zero second neighbor hopping amplitude $|t'|/t=0.2$.

Recently, Shastry~\cite{shastry_1,shastry_2,shastry_3} predicted a
low to intermediate temperature enhancement of the thermopower via
a high temperature expansion of the high frequency limit $S^*(T)$ for
the geometrically frustrated two-dimensional triangular lattice
$t$-$J$ model.  In that case, changing the sign of the hopping was
found to enhance the thermopower at intermediate temperatures.
Haerter et. al.\cite{cw_prl} and Ref.~\onlinecite{peterson}
more thoroughly investigated that particular case.

In one-dimension it is similarly expected that an enhancement of
the thermopower will occur when a second-neighbor hop is added to
the kinetic energy which
frustrates the lattice and destroys integrability.

Fig.~\ref{sstar_t-02}(a)-(d) show $S^*(T)$ versus temperature and filling
for values of $t'/t=-0.2$ (Fig.~\ref{sstar_t-02}(a)-(c))
and $t'/t=0.2$ (Fig.~\ref{sstar_t-02}(d)).  For this case
we do not plot the full frequency dependence but remark that it is
similar qualitatively and quantitatively to the $t'=0$ case\cite{t02-note}.

\begin{figure}[t]
\includegraphics[width=7.cm]{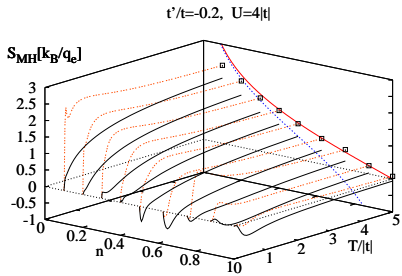}
\caption{(color online) $S_{MH}(T)$ (black line) as a function of filling $n$ and temperature $T$.
The orange dotted lines are the
infinite frequency limit $S^*(T)$ for comparison.  Projected
onto the $T=5|t|$ plane are the MH limits for both the finite (red line) and
infinite $U$ (blue dotted line) cases.}
\label{smh_t-02}
\end{figure}

When $t>0$ and $t'<0$ the thermopower is
enhanced at low temperatures.  The enhancement is seen to
arise from almost purely the transport term of the thermopower (first
term in Eq.~\ref{therm1}).  Fig.~\ref{smh_t-02} shows $S^*(T)$ and
the MH term $S_{MH}(T)$ for $t'/t=-0.2$ and $U=4|t|$ as a representative
example.  Similar to the results displayed in Fig.~\ref{smh_t0} the low temperature
thermopower is dominated by the transport term and in the case of $t'/t=-0.2$ that
domination is even more pronounced as it produces an enhancement peak.

For small fillings the enhancement has weak $U$ dependence although the non-interacting
case seems to be tending to diverge at very low temperatures ($n=0.1$ and $n=0.2$
in Fig.~\ref{sstar_t-02}(a)) that is most certainly a finite size
artifact.  At fillings above $n=0.5$ there are interaction
effects which are visible at low temperatures.  For the frustrated
case discussed here the thermopower still mostly achieves its
MH limit by $T=5|t|$ as expected.

For the situation when $t$ and $t'$ are both positive
there is a suppression of the thermopower at low temperatures,
 c.f. Fig.~\ref{sstar_t-02}(d), hence the opposite effect.  Only
the weakly correlated $U=4|t|$ case is plotted to illustrate this.

For both signs of $t'$ at half filling ($n=1$) the thermopower is no longer
identically zero since the addition of a
non-zero $t'$ destroys the particle-hole symmetry, however,
the thermopower remains quite small at this density.

\subsection{$t$-$V$ model}\label{sec-tV}

\begin{figure}[t]
\begin{center}
\mbox{\bf (a)}
\includegraphics[width=7.cm]{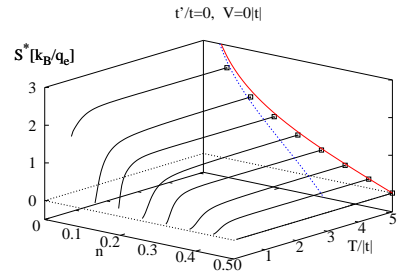}\\
\mbox{\bf (b)}
\includegraphics[width=7.cm]{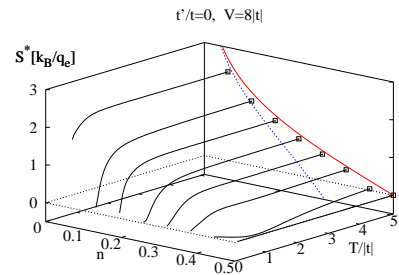}\\
\mbox{\bf (c)}
\includegraphics[width=7.cm]{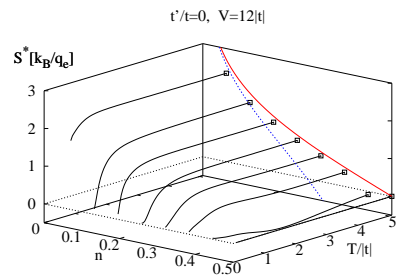}
\end{center}
\caption{(color online) $S^*(T)$ (black line) as a function of filling $n$ and temperature $T$.
For (b) and (c) the orange dotted lines are the
DC limit $S(0,T)$ of the full thermopower for comparison.  Projected
onto the $T=5|t|$ plane are the MH limits for both the finite (red line) and
infinite $V$ (blue dotted lines) situations.}
\label{sstar_t0_V}
\end{figure}

\begin{figure*}[t]
\centering
\begin{tabular}{cc}
\mbox{\bf (a)}
\includegraphics[width=7.cm]{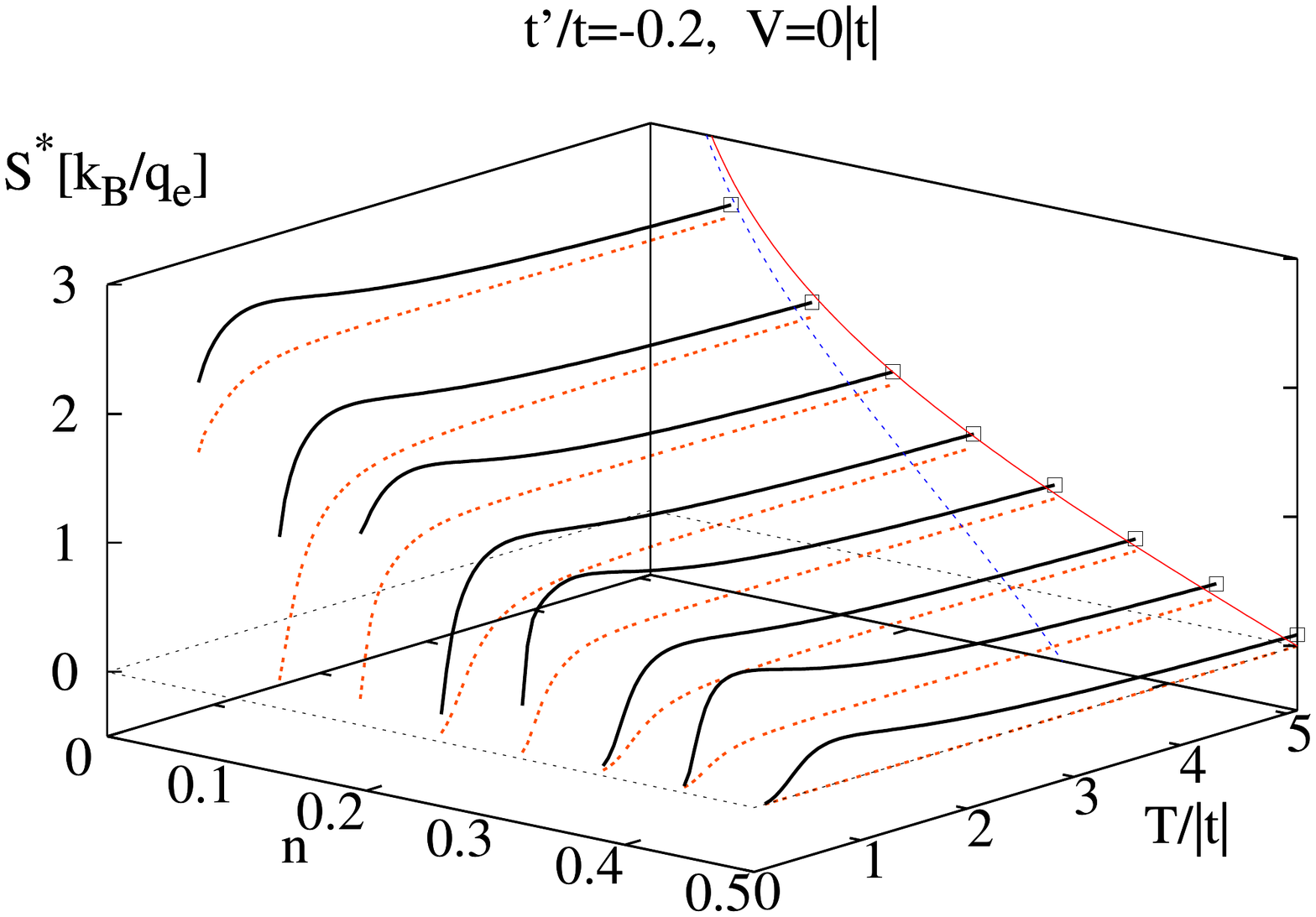}&
\mbox{\bf (b)}
\includegraphics[width=7.cm]{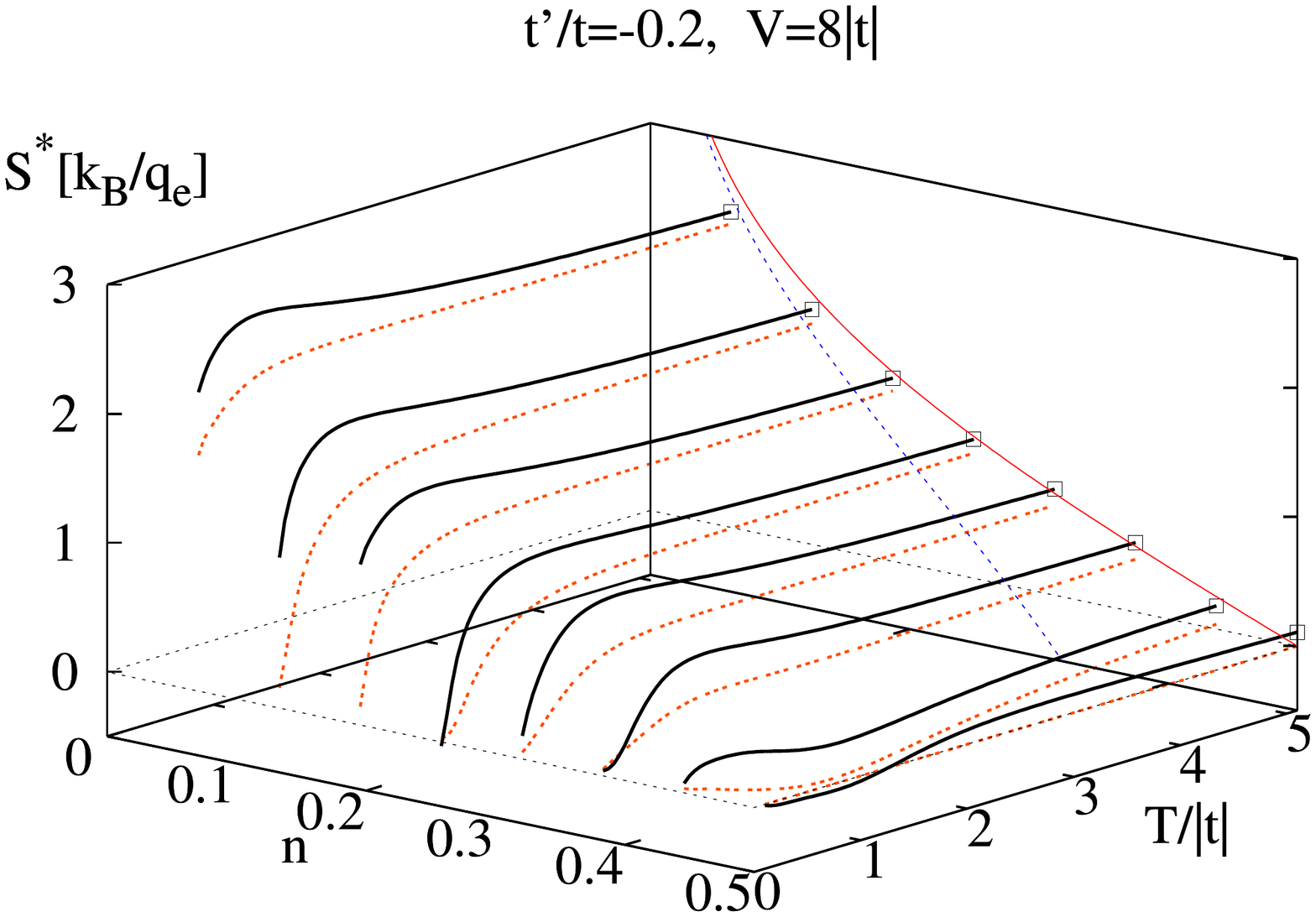}\\
\mbox{\bf (c)}
\includegraphics[width=7.cm]{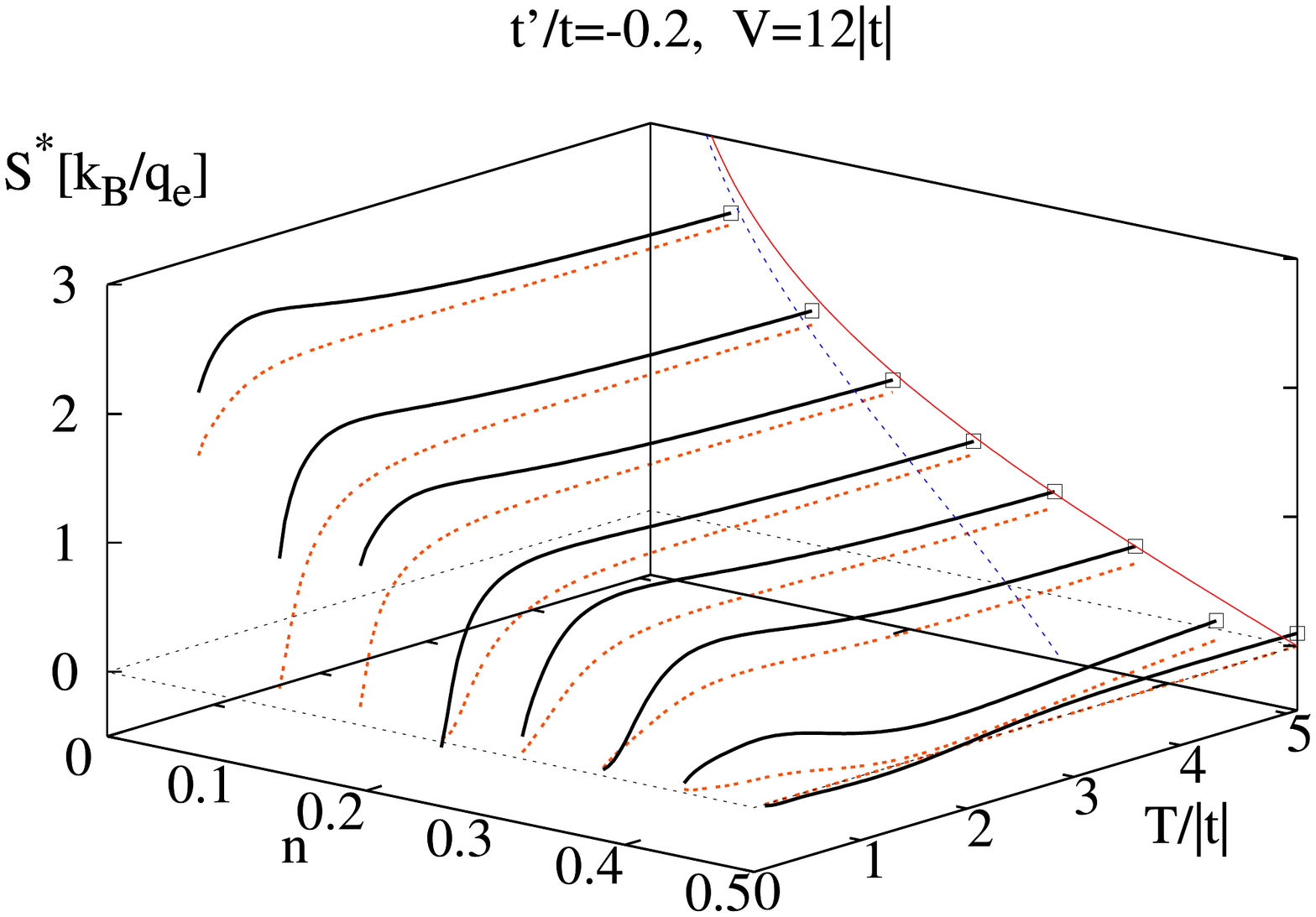}&
\mbox{\bf (d)}
\includegraphics[width=7.cm]{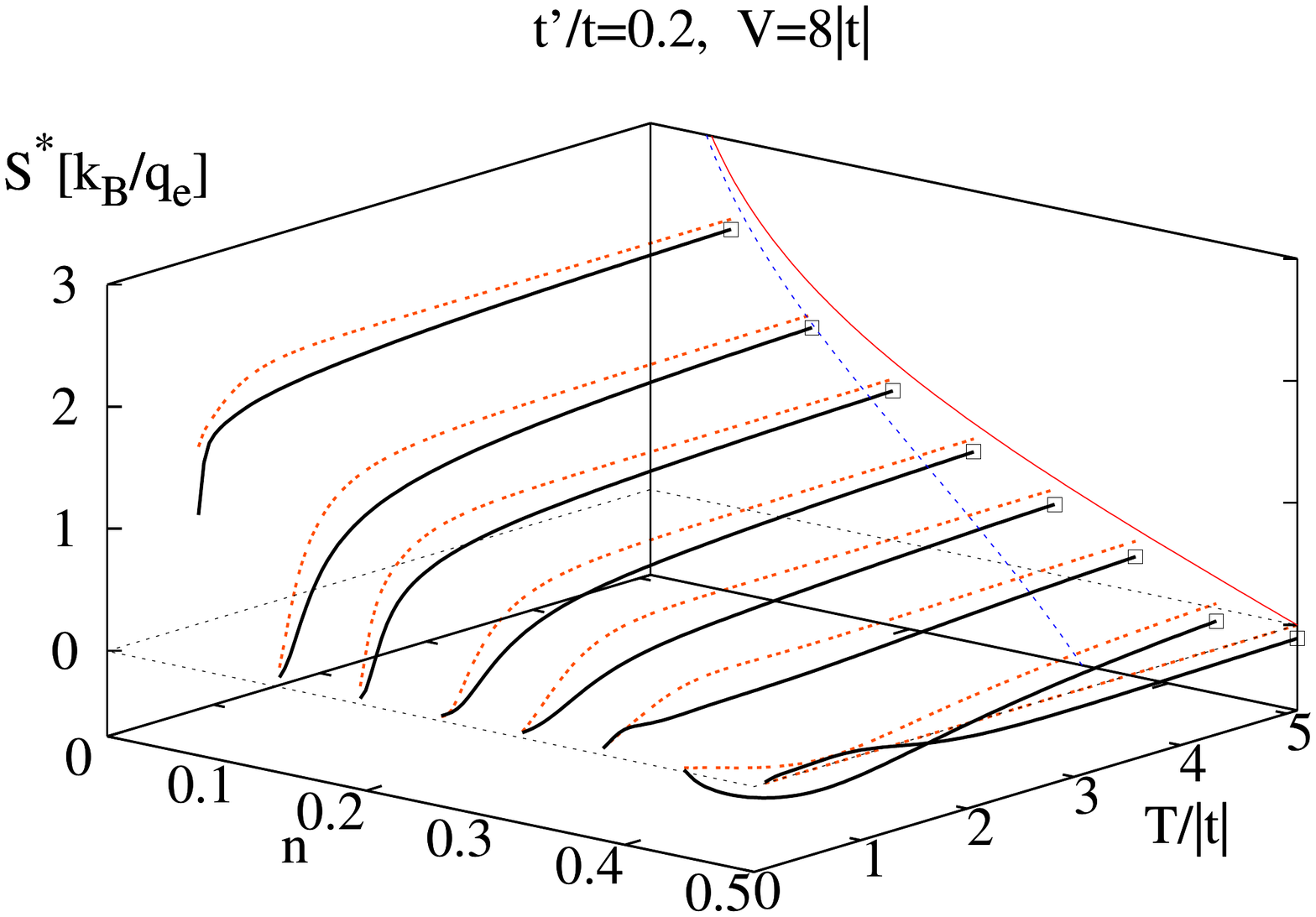}
\end{tabular}
\caption{(color online) $S^*(T)$ as a function of filling $n$ and temperature $T$ for
(a) $V=0$, (b) $V=8|t|$, and (c) $V=12|t|$.  The
full $S(\omega,T)$ compares similarly to $S^*(T)$ as for the $t'=0$ case, c.f. Fig.~\ref{s-sstar_t0}.
The red curve is $S^*(T)$ for $t'=0$ to facilitate an easy comparison.  As
an example of the other sign of $t'$, namely, $t'/t=0.2$, (d) shows the expected
reduction in $S^*(T)$ for $V=8|t|$.  The MH limits are projected unto the $T=5|t|$ plane
for finite $V$ (red line) and infinite $V$ (blue dotted line).}
\label{sstar_t-02_V}
\end{figure*}

Most of the considerations of the thermopower for the Hubbard
model also apply to the $t$-$V$ model. In particular this model
too has particle hole symmetry at half-filling ($n=0.5$ instead of
$n=1$ as for the Hubbard model, since the spin degree of
freedom is absent) and thus the thermopower is identically zero at all
temperatures. The thermopower is once again divergent in the
band-insulator limits ($n=0$ and $n=1$) and also in the vicinity
of the Mott insulator ($n=0.5$) in the presence of strong
correlations (large $V$). In the high temperature limit ($k_BT \gg
t$), the thermopower is pinned by the entropy as in the Hubbard
model. The MH limit can be
calculated in a straightforward manner in the $V=0$ and $V=\infty$
regimes from the expressions already derived for the extended
Hubbard model (on-site $U$ and nearest-neighbor
$V$)~\cite{chaikin-beni}. The $t$-$V$ model corresponds to setting
$U=\infty$ and removing the spin degeneracy. We thus obtain
\begin{eqnarray}
\lim_{T\rightarrow\infty}S_{MH}(T) = \left\{ \begin{array}{l l}
  \frac{k_B}{q_e}\ln \left(\frac{1-n}{n}\right)& \quad \mbox{(i)}\\
  \frac{k_B}{q_e}\ln \left(\frac{(1-2n)^2}{n(1-n)}\right) & \quad \mbox{(ii)}
\end{array}\;,
\label{mh-tv}
\right.
\end{eqnarray}
where Eq.~\ref{mh-tv}(i) corresponds to finite $V$ and $0\leq n\leq 1$
while Eq.~\ref{mh-tv}(ii) corresponds to infinite $V$ and $0\leq n \leq 0.5$.
As mentioned previously, the expressions in Eq.~\ref{mh-tv} were first
considered in Ref.~\onlinecite{chaikin-beni} with the 
the second line (the infinite $V$ limit) being the third equation 
of Table I in Ref.~\onlinecite{chaikin-beni} with the spin 
degeneracy removed.

An important distinction between the Hubbard model and the $t$-$V$
model is that the energy current operator commutes with the
Hamiltonian in the latter. From the Kubo formula, this implies
that in Eqn.~\ref{therm1}, $S_{tr}(\omega,T)=0$ for all $\omega
\neq 0$ and $S(\omega, T) = S_{MH}(T)$ for all $\omega \neq 0$.
Since $S^*(T) = S(\omega \rightarrow \infty, T)$, this leads to 
$S^*(T) = S_{MH}(T)$ at all temperatures and filling in this
model. From Eqn.~\ref{sstar}, we have 
that $\langle\hat\Phi_{xx} \rangle=-\mu \langle \hat\tau_{xx}\rangle$, a
fact that has been verified numerically. A further consequence is
that $\kappa(\omega,T)/T\sigma(\omega,T)=S(\omega,T)^2$ for all
$\omega \neq 0$ and consequently $L(\omega > 0,T)=0$ implying
$Z(\omega > 0,T)T = \infty$ in this system. Physically, the energy
current commuting with the Hamiltonian means that the system is
unable to transport any heat at finite frequency in the presence
of a temperature or potential gradient without transporting
charge. In zero current conditions where there is no charge
transport, there is no heat transport as well and the thermal
conductivity is zero. This causes the Lorenz number to be zero and
the FOM to be infinite. All the above considerations rely on the
fact that the energy current commutes with the Hamiltonian. This
is no longer true with $t' \neq 0$ and all the quantities
mentioned above will have have non-zero values at $\omega \neq 0$.

\subsubsection{$t$-$V$ with $t'=0$}

The behavior of the thermopower for the $t$-$V$ model is quite similar
to that of the Hubbard model even though the thermopower has
no transport term for $t'=0$ and is, therefore, simply $S_{MH}(T)$.
Plotted in Fig.~\ref{sstar_t0_V}(a)-(c) is the thermopower $S^*(T)$
for three values of the interaction strength, $V=0$, $V=8|t|$,
and $V=12|t|$.  At low densities the thermopower very rapidly
rises to nearly its full MH limit by $T\approx 2|t|$.  The initial
(mostly) linear slope is reduced as the filling is increased.  The effects
of interactions are not readily seen until the somewhat
large filling of $n=0.4375$ where the thermopower is markedly reduced
for all temperatures shown.  Eventually, as the temperature is raised, the
interaction effects are washed out as the thermopower approaches its MH limit.

Compared to the Hubbard model, however, the interaction effects appear
weaker (perhaps due to the absence of the transport term)
and effect only the highest fillings studied (other than the half filled
case $n=0.5$ which is pinned at zero due to particle-hole symmetry).

\subsubsection{$t$-$V$ with $t'\neq0$}
Adding a second-neighbor $t'$ hopping term has an effect very
similar to the one in the Hubbard model again destroying the
integrability and the particle-hole symmetry of the model.  A
result of the latter is that the thermopower is no longer
identically zero at half-filling. As in the Hubbard model, a
more interesting aspect of introducing a second-neighbor hopping
is that it produces frustration depending on its sign.

The thermopower is plotted as a
function of temperature and filling for $V=0$, $V=8|t|$, and
$V=12|t|$ for the case of $t'/t=-0.2$ in Fig~\ref{sstar_t-02_V}(a)-(c),
and for $V=8|t|$ for the case of $t'/t=0.2$ in Fig~\ref{sstar_t-02_V}(d).

A positive sign of $t'$ reduces the value of $S^*(T)$ compared to
$t'=0$, while a negative sign enhances it. $S^*(T)$ starts out being
zero at $T=0$ and approaches the Mott-Heikes limit as $T
\rightarrow \infty$ independent of the value of $t'$. Since $S^*(T)$
for $t'=0$ is a monotonically increasing function of $T$, it stands
to reason that $S^*(T)$ reaches a maximum at some value of $T$ for
$t' <0$ and decreases towards the Mott-Heikes limit. This is
indeed what is seen in our calculations, as demonstrated in
Fig.~\ref{sstar_t-02_V}, similar to the situation 
of the Hubbard model.  This
is very interesting since it affords
the possibility of thermopower enhancement through frustration.

\begin{figure}[t]
\begin{center}
\mbox{\bf (a)}
\includegraphics[width=7.cm]{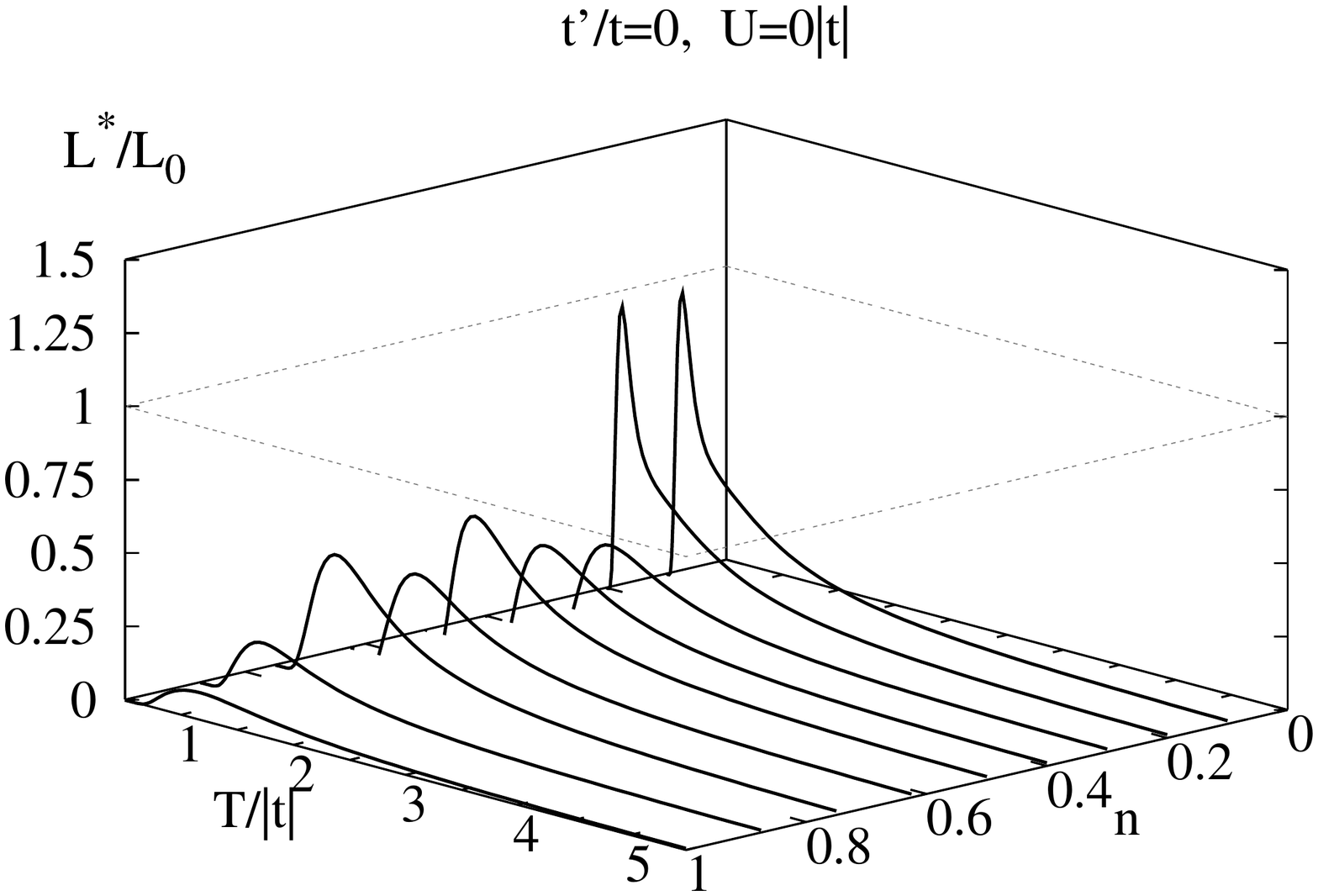}\\
\mbox{\bf (b)}
\includegraphics[width=7.cm]{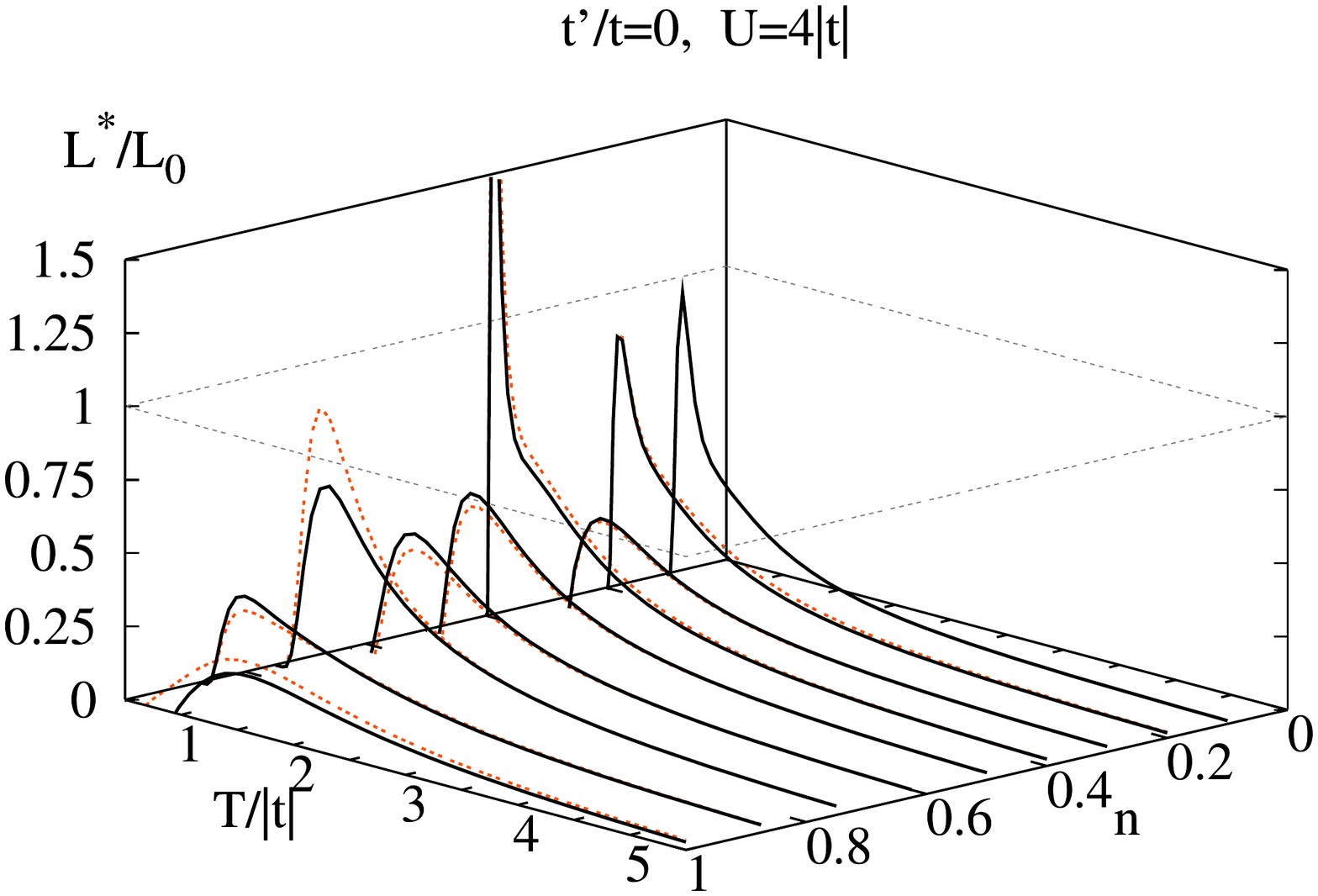}\\
\mbox{\bf (c)}
\includegraphics[width=7.cm]{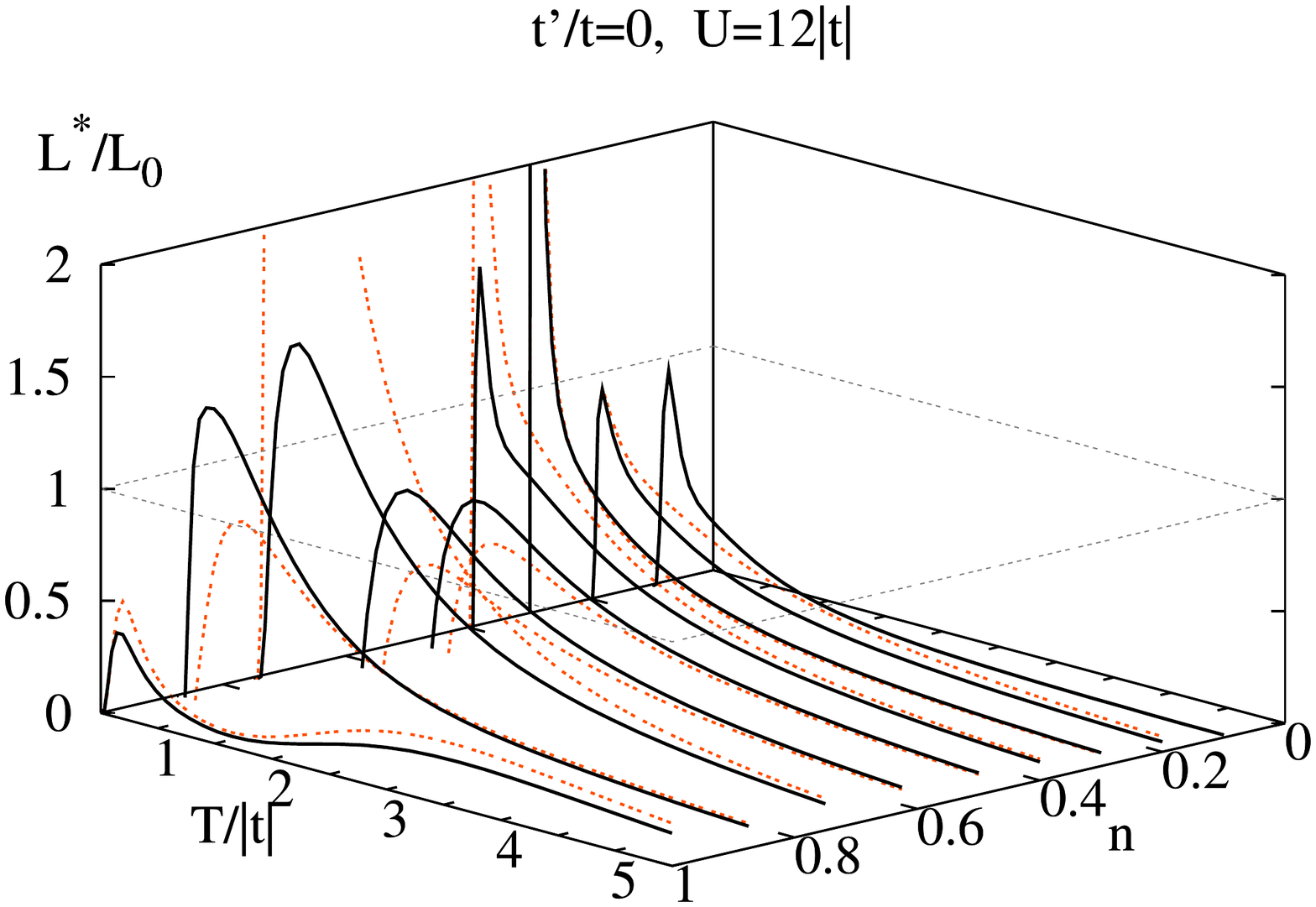}
\end{center}
\caption{(color online) $L^*(T)/L_0$ as a function of filing $n$ and
temperature $T$ (black line). The
orange dotted line is the DC limit $L(0,T)$ for comparison.  Note that for
$U=0$ (a) there is no frequency dependence of $L(\omega,T)$ and $L^*(T)=L(0,T)$.
The full $L(\omega,T)$ (not shown) compares similarly to the
frequency dependence of the thermopower for $U=4|t|$
and $12|t|$ in Fig.~\ref{s-sstar_t0}.}
\label{lstar_t0}
\end{figure}

\begin{figure*}[t]
\centering
\begin{tabular}{cc}
\mbox{\bf (a)}
\includegraphics[width=7.cm]{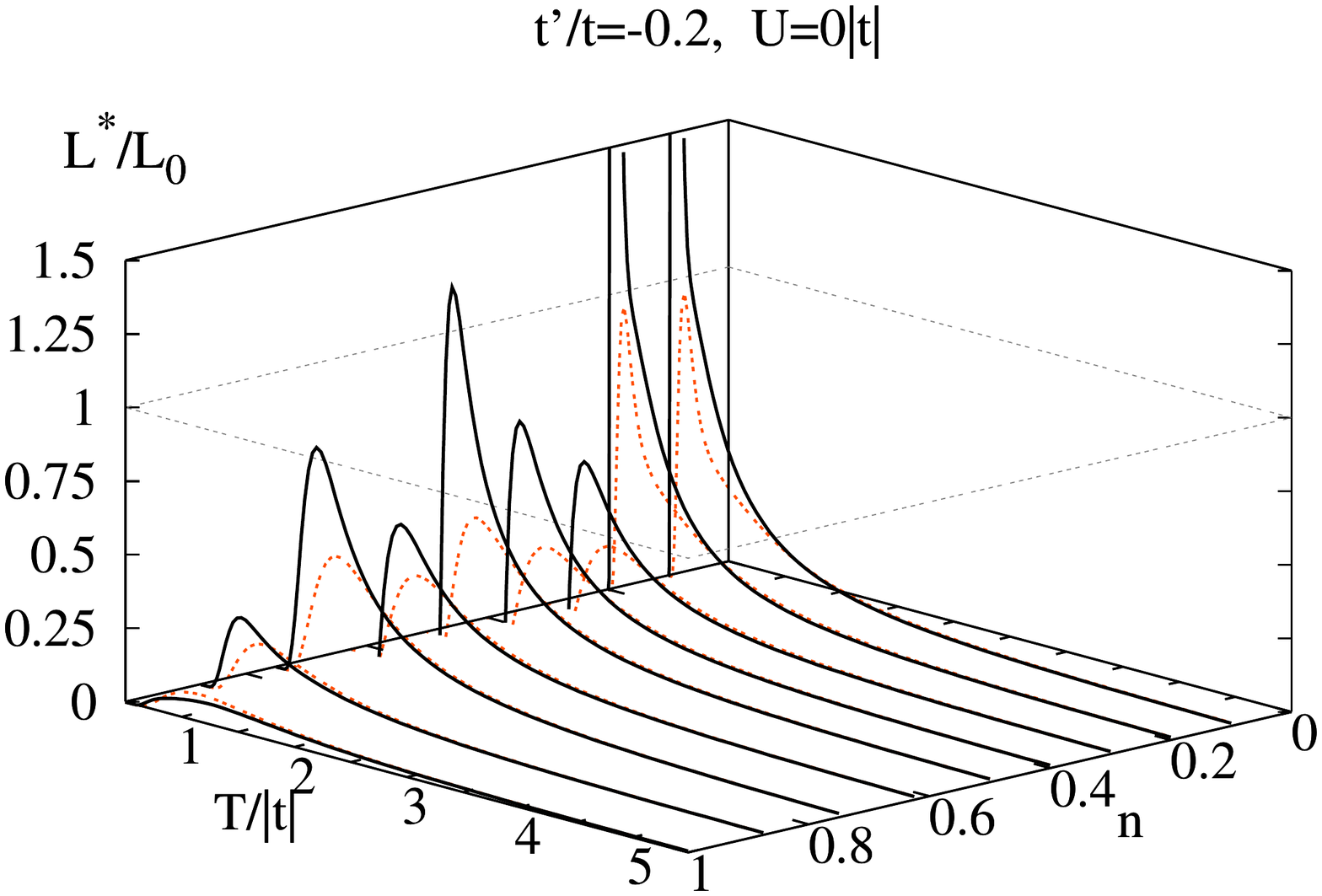}&
\mbox{\bf (b)}
\includegraphics[width=7.cm]{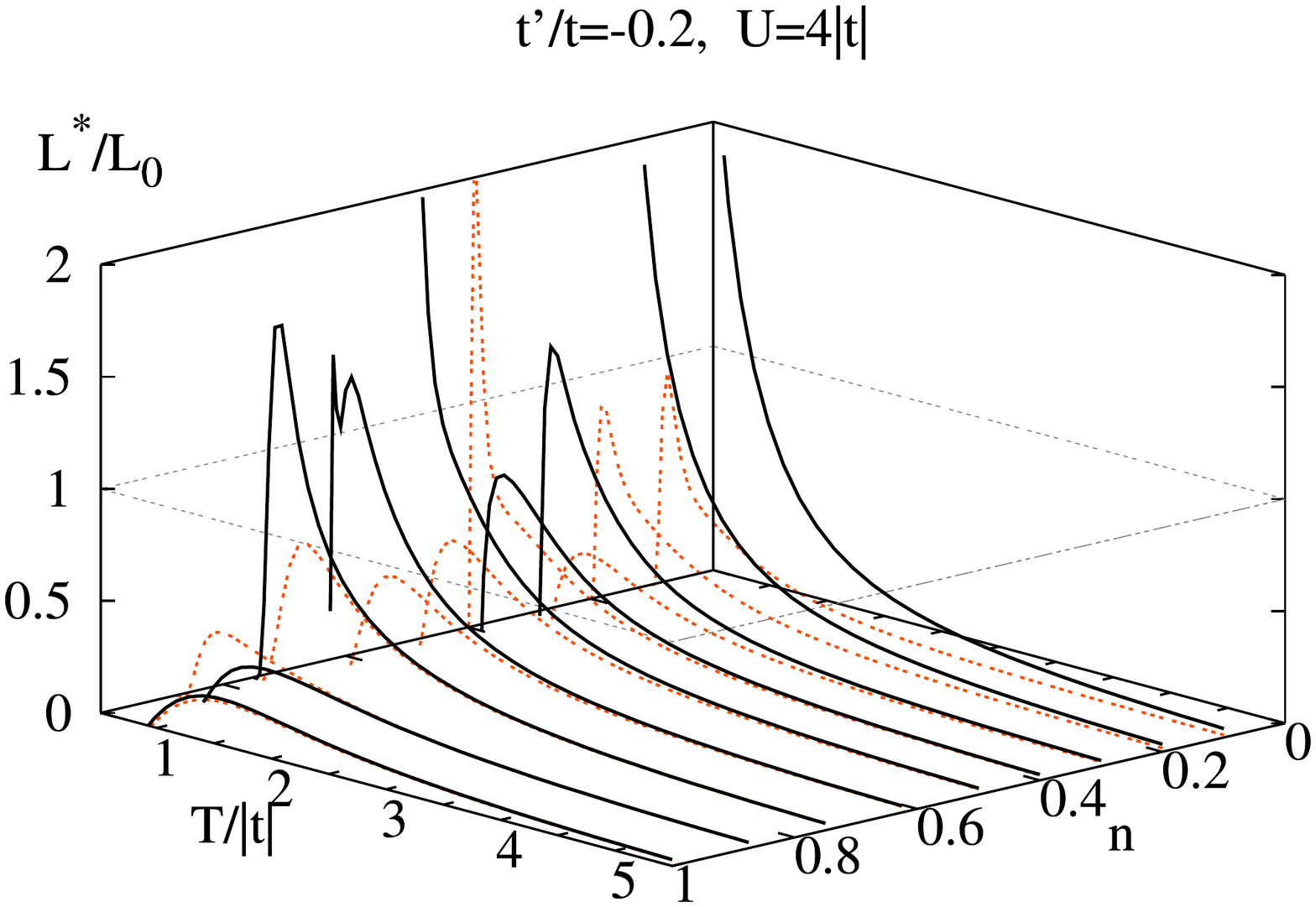}\\
\mbox{\bf (c)}
\includegraphics[width=7.cm]{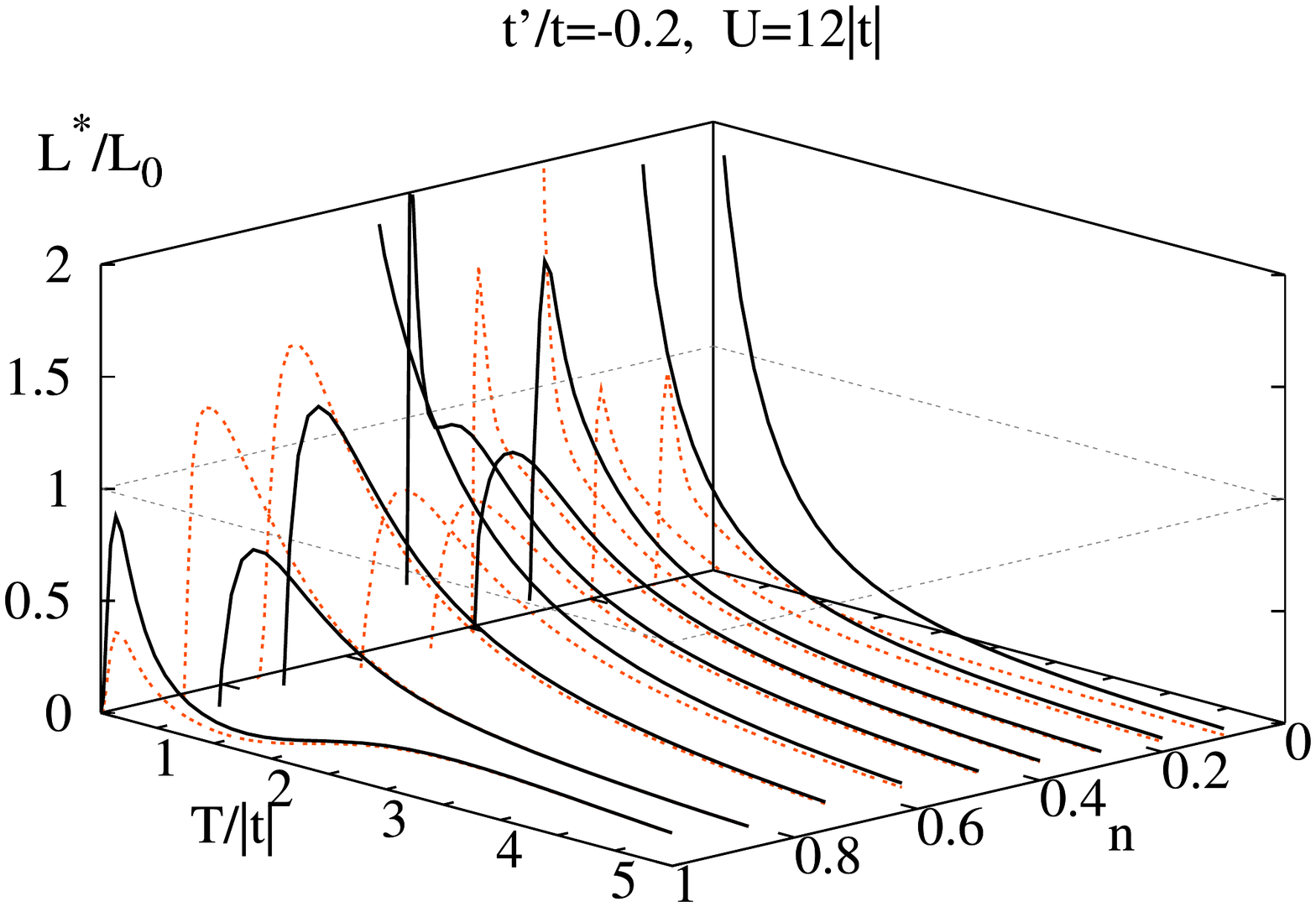}&
\mbox{\bf (d)}
\includegraphics[width=7.cm]{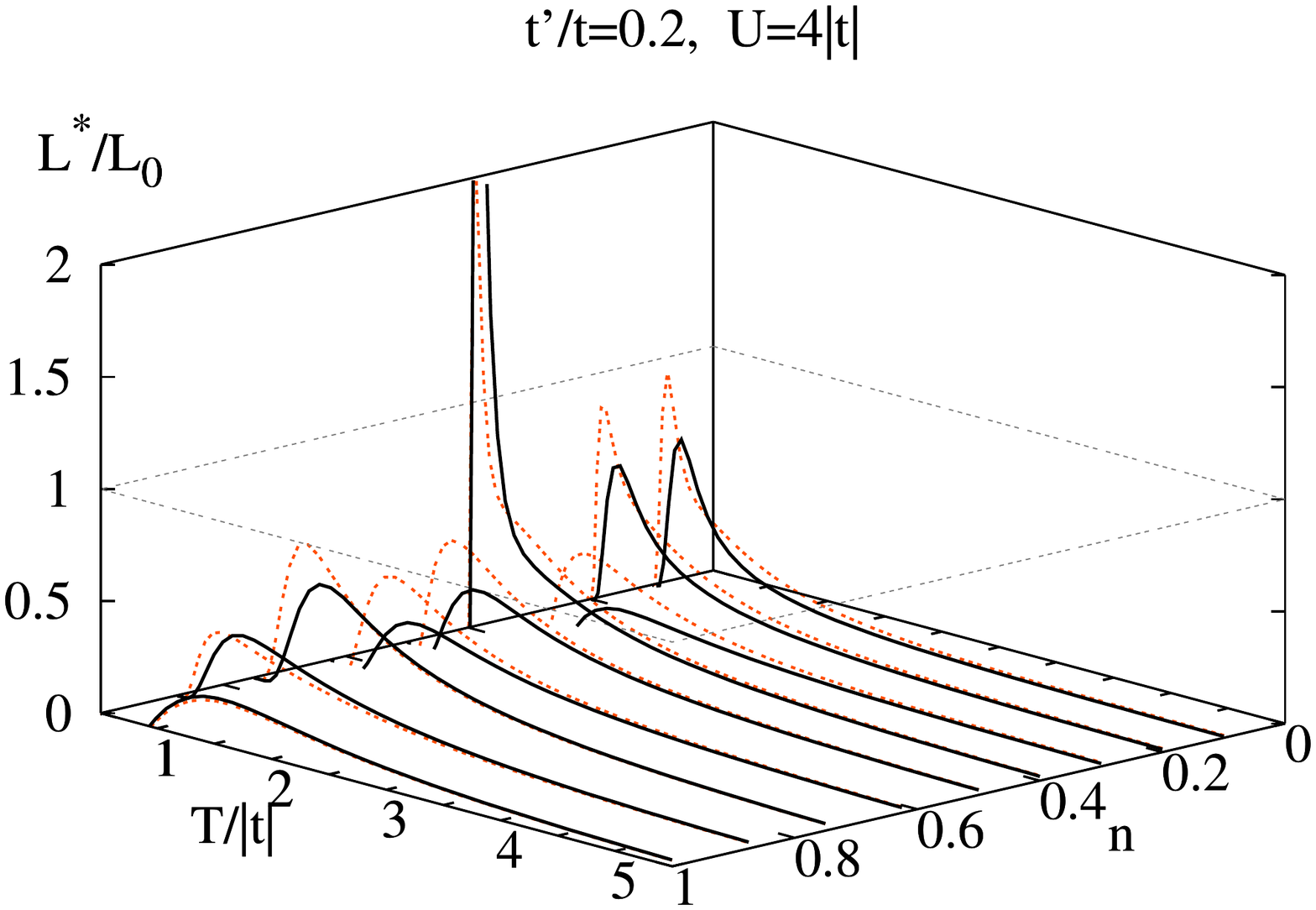}
\end{tabular}
\caption{(color online) $L^*(T)/L_0$ as a function of filling $n$ and temperature $T$ for
(a) $U=0$, (b) $U=4|t|$, and (c) $U=12|t|$.  The
full $L(\omega,T)$ compares similarly to frequency dependence of the thermopower.
The orange dotted curve is the $t'=0$ case to facilitate an easy comparison.  (d)
is an example of the other sign of $t'$, namely, $t'/t=0.2$.}
\label{lstar_t-02}
\end{figure*}

Even though we do not explicitly calculate $S(\omega,T)$ for the $t$-$V$ model
when $t'\neq 0$,
we comment that the introduction of $t'$ causes the energy current operator
not to commute with Hamiltonian. Consequently $S(\omega,T)$ is
now no longer independent of $\omega $ and
approaches $S^*(T)$ as $\omega \rightarrow \infty$ like in the
Hubbard model.  Presumably, the full frequency dependence is similar to
that seen in the Hubbard model but has not
been calculated explicitly.

Although not shown for the $t$-$V$ model,
we point out that similar to the Hubbard model, the thermopower
enhancement peak arises from the transport term of the thermopower
almost exclusively as discussed in
Sec.~\ref{hub-t-02} and shown in Fig.~\ref{smh_t-02}.

\section{Lorenz number}\label{sec-l}

The Lorenz number (Eq.~\ref{lorenz}) is an important quantity as it measures
the ratio of the thermal to electrical conductivity.  Further, it is a key
ingredient to the FOM (containing both the thermopower and Lorenz number)
which measures the efficiency of a thermoelectric material.  Note that 
there is usually a contribution to the Lorenz number coming from lattice
vibrations (phonons).  In this work, however, we only consider the
electronic contributions to the Lorenz number.

As noted in Ref.~\onlinecite{peterson} the chemical potential is 
absent from Eq.~\ref{lorenz} and the Lorenz number can be understood completely
within the canonical ensemble.  Evidently, it is determined
through electron transport alone.  At zero temperature it is well known that the
Lorenz number is equal to the constant $L_0=(\pi k_B/\sqrt{3}q_e)^2$ which
is just the Wiedemann-Franz law\cite{am,ziman}.

As discussed in Ref.~\onlinecite{peterson}, for
non-interacting systems, the limit $\lim_{T\rightarrow0}L(\omega,T)=L_0$
comes from two effects.  Similar to the thermopower, we attempt
to limit the finite sized system induced divergences of the
Lorenz number by forcing the subtle balances necessary to ensure
a finite Lorenz number as $T\rightarrow 0$.  The exact method
used here is the same as in Ref.~\onlinecite{peterson} and
not repeated.  Ultimately, we are able to control
the divergences to a large degree, however, we are
unable to answer the very interesting question of
whether the value of $L_0$ is a universal constant independent of
electron interactions.

Below we present $L^*(T)$ as a function of temperature
and filling.  The frequency dependence of $L(\omega,T)$ (not shown here)
is comparable to $S(\omega,T)$ in that it is generally weak
with a feature near $\omega=U$.

\subsection{Hubbard with $t'=0$}

Fig.~\ref{lstar_t0}(a)-(c) shows the ``normalized'' high
frequency Lorenz number $L^*(T)/L_0$
as a function of temperature
and filling for the non-interacting, the weakly coupled, and the strongly coupled
situations for the case of $t'=0$.  In Fig.~\ref{lstar_t0}(b)-(c)
the DC limit of the full frequency dependent Lorenz
number is also plotted, i.e., $L(0,T)/L_0$.

For the non-interacting case (Fig.~\ref{lstar_t0}(a)) $L^*(T)$
is suppressed at low temperatures as the filling increases towards
half filling.  For a thermodynamically large system the Lorenz number
starts at $L_0$ and quickly decreases as a function of temperature
similar to what is shown here.  In the weakly
coupled regime (Fig.~\ref{lstar_t0}(b)) $L^*(T)$
is very similar to the non-interacting case for fillings below
approximately $n=0.4$.  For $n\geq0.5$, however, it is
enhanced at low temperatures.  The DC limit is quite similar
to the infinite frequency limit showing the usefulness and accuracy of
$L^*(T)$ in the weakly coupled regime.  When the interactions
are strong (Fig.~\ref{lstar_t0}(c)) there
is a much stronger enhancement of $L^*(T)$ for $n\geq 0.3$
and the enhancement persists to higher temperatures.  The
DC limit in this case is not as similar to $L^*(T)$ as it is for the weakly coupled
regime.

\subsection{Hubbard with $t'\neq 0$}

For non-zero $t'$ we plot $L^*(T)/L_0$ as a function of filling
and temperature along with $L^*(T)/L_0$ for the $t'=0$
to ease comparison in Fig.~\ref{lstar_t-02}(a)-(d).

For the same sign of $t'<0$ that produced the thermopower
enhancement the Lorenz number in Fig.~\ref{lstar_t-02}(a)-(c) is also
enhanced compared to the $t'=0$ situation at
low temperatures.  This enhancement is also more pronounced
for low than for high fillings and increases with
increasing interaction strength $U$, especially close to half filling.

When $t'>0$ in the weakly coupled regime (Fig.~\ref{lstar_t-02}(d)) the
Lorenz number is very slightly suppressed compared to the $t'=0$ situation
as one would expect from the $t'<0$ results and the previous
investigation of the thermopower.

Generally, for both $t'=0$ and $t'\neq 0$ the Lorenz number is
appreciably below $L_0$ for all temperatures above approximately
$T\sim 1.5|t|$.

\section{Figure of merit}\label{sec-fom}

\begin{figure}[t]
\begin{center}
\mbox{\bf (a)}
\includegraphics[width=7.cm]{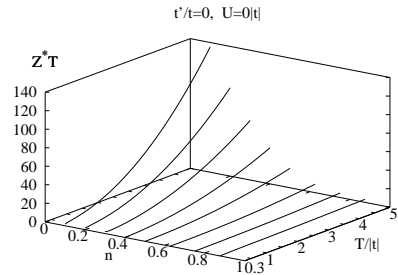}\\
\mbox{\bf (b)}
\includegraphics[width=7.cm]{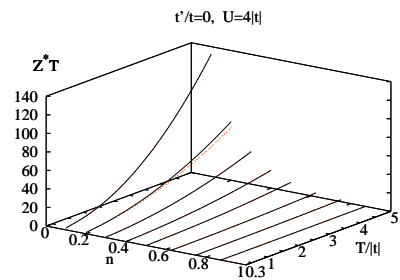}\\
\mbox{\bf (c)}
\includegraphics[width=7.cm]{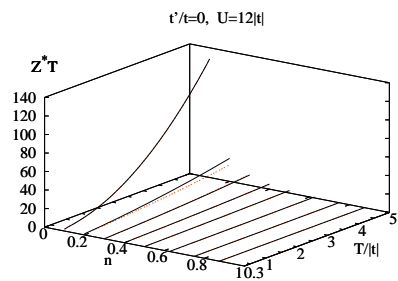}
\end{center}
\caption{(color online) $Z^*(T)T$ as a function of filling $n$ and temperature $T$ as
the solid black line for $U=0$ (a), $U=4|t|$ (b),
and $U=12|t|$ (c). The orange dotted curve is the DC limit of the
full $Z(\omega,T)T$ for comparison, i.e., $Z(0,T)T$.  Note that for $U=0$ (a) there
is no frequency dependence.}
\label{zstart_t0}
\end{figure}

\begin{figure*}[t]
\centering
\begin{tabular}{cc}
\mbox{\bf (a)}
\includegraphics[width=7.cm]{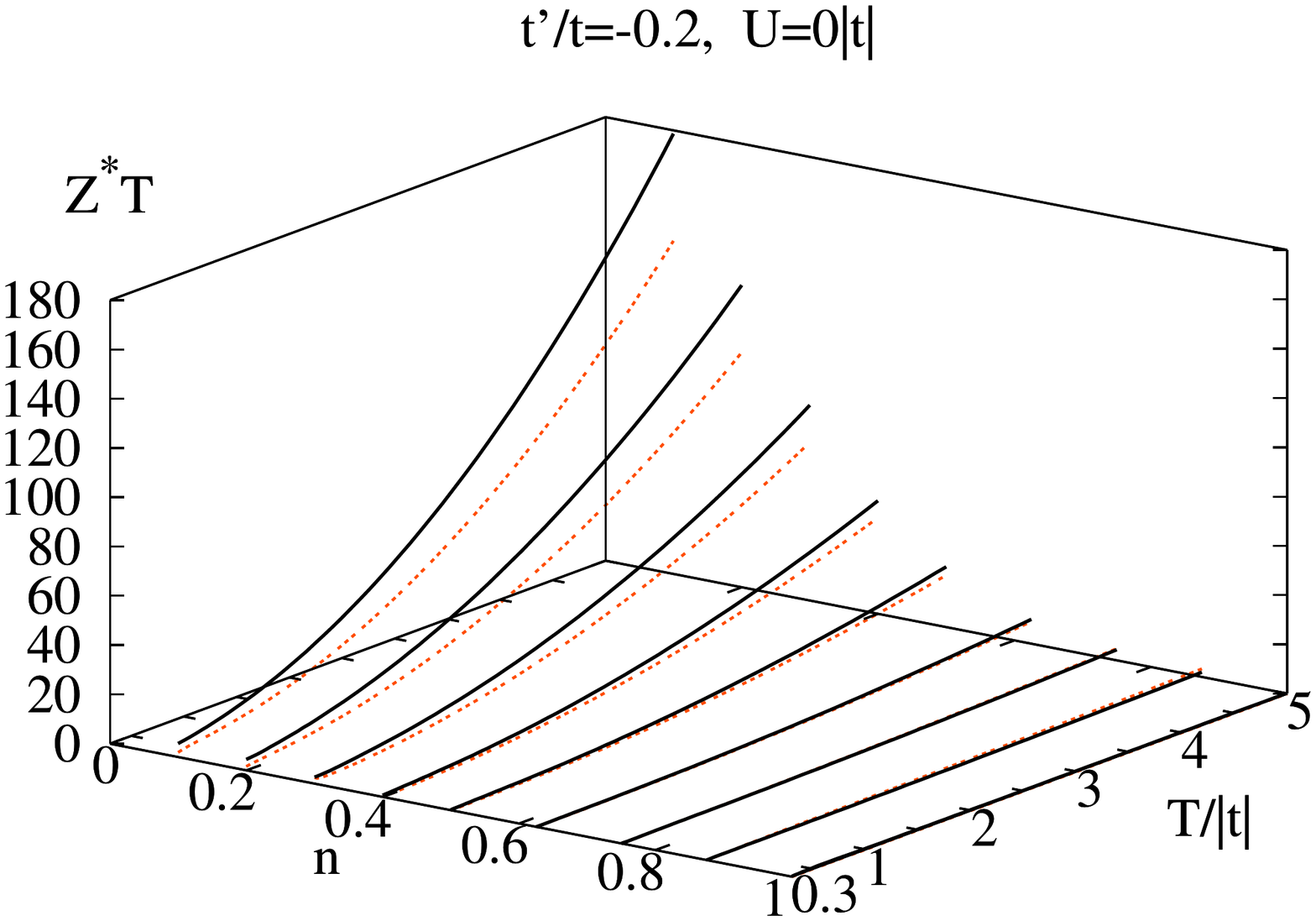}&
\mbox{\bf (b)}
\includegraphics[width=7.cm]{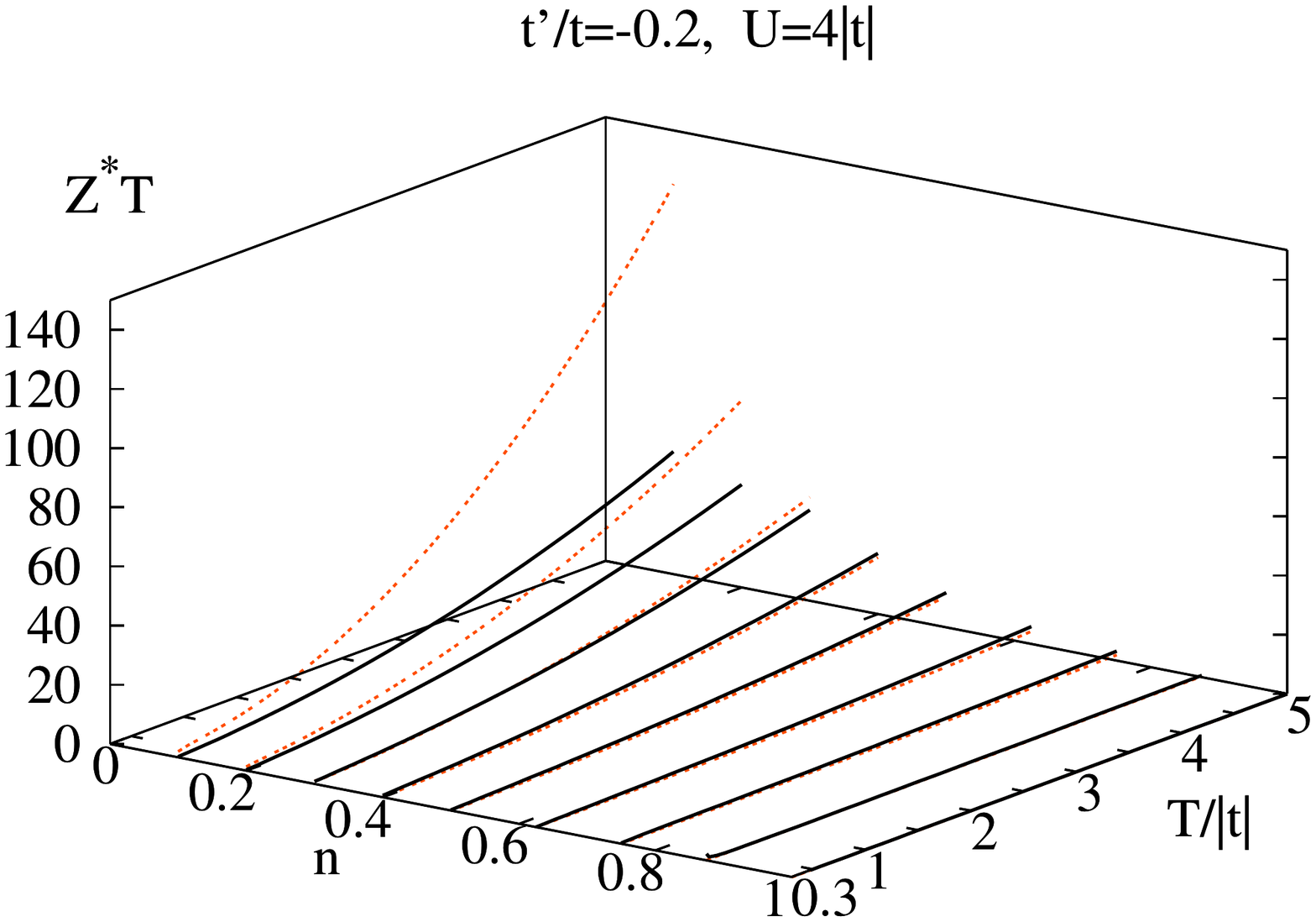}\\
\mbox{\bf (c)}
\includegraphics[width=7.cm]{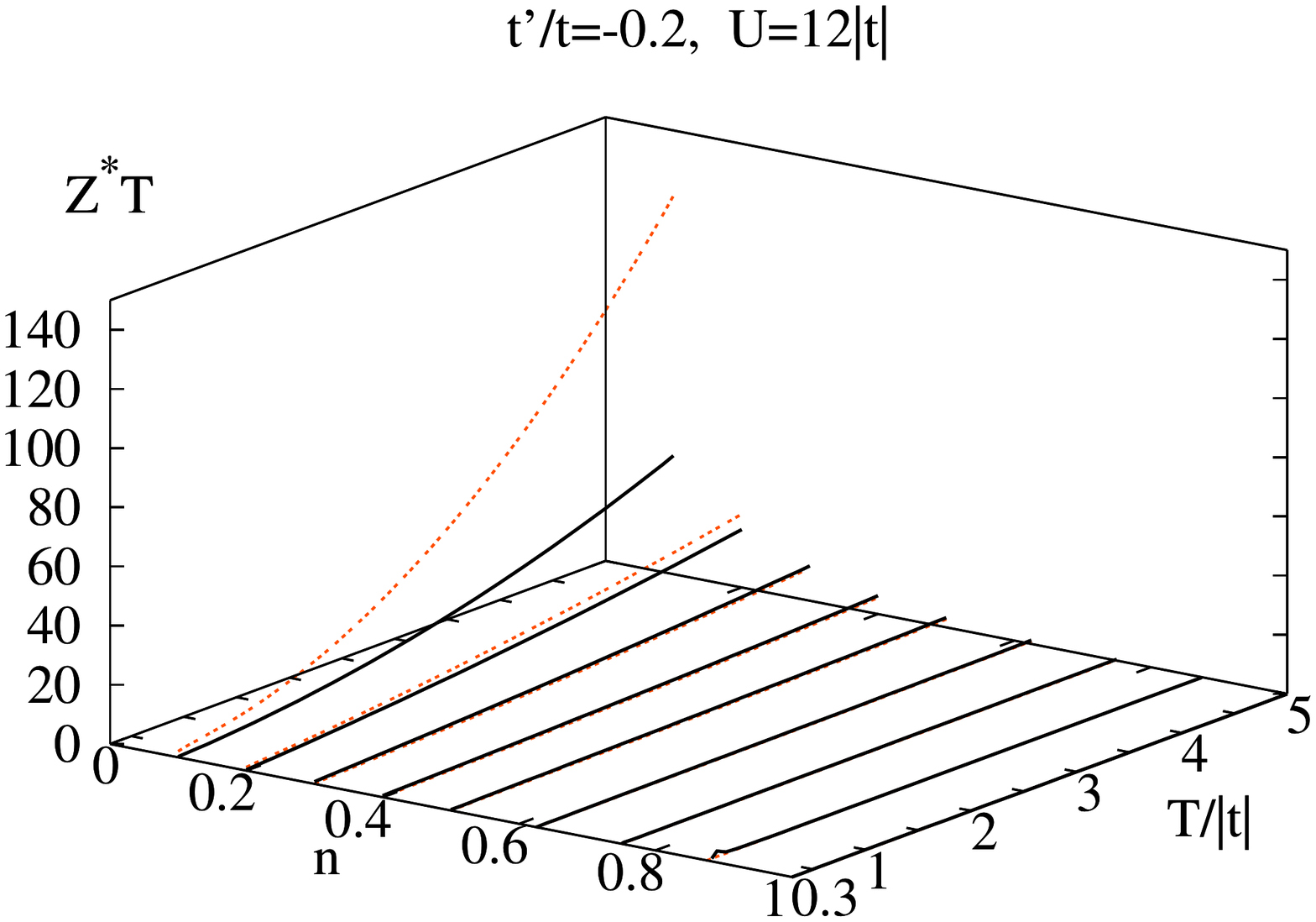}&
\mbox{\bf (d)}
\includegraphics[width=7.cm]{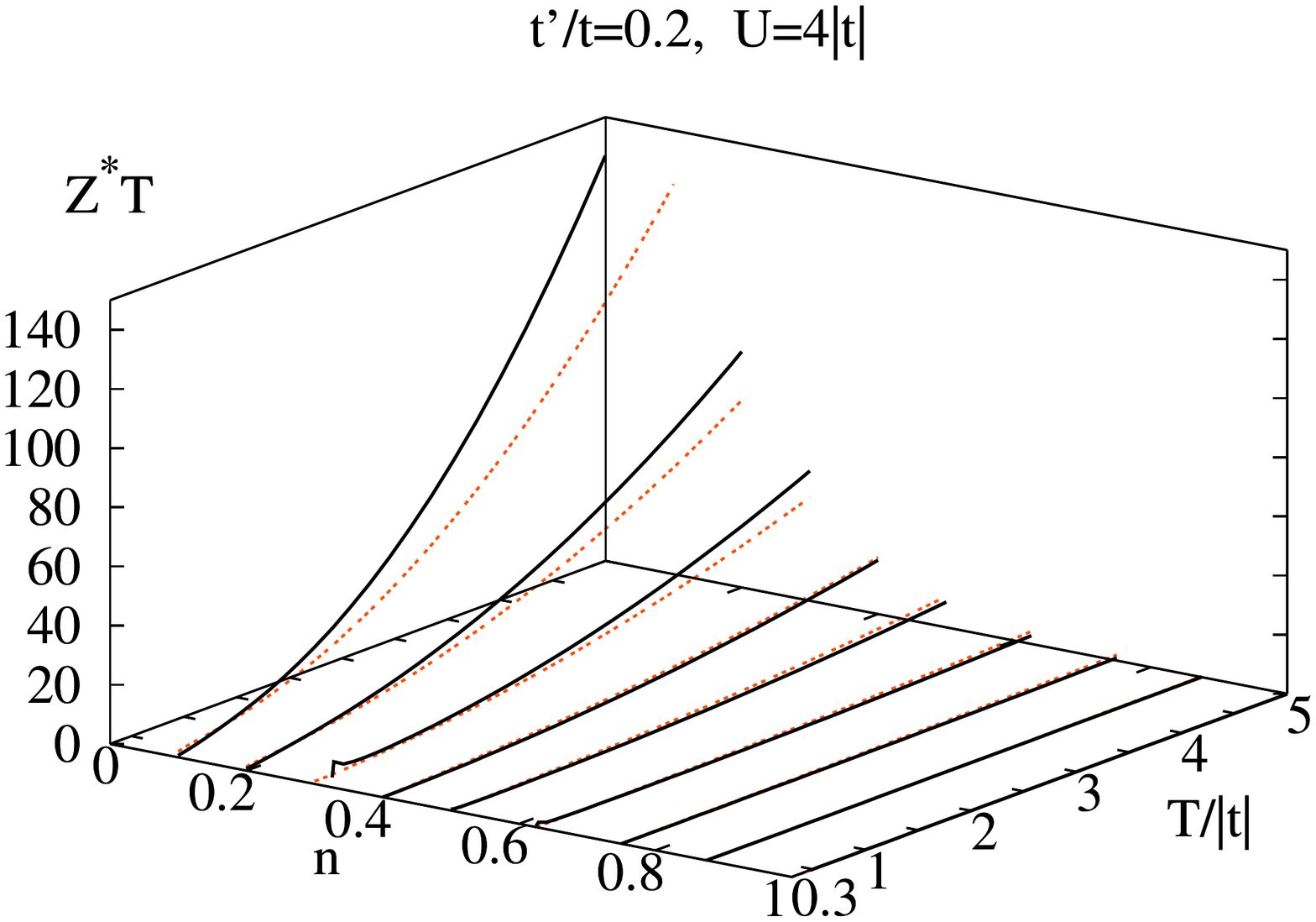}
\end{tabular}
\caption{(color online) $Z^*(T)T$ as a function of filling $n$ and temperature $T$ for
(a) $U=0$, (b) $U=4|t|$, and (c) $U=12|t|$.  The
full frequency dependent $Z(\omega,T)T$ compares similarly to the
frequency dependence of the
thermopower.  The orange dotted curve is for $t'=0$ to facilitate an easy
comparison.  For an example of the other sign of $t'$ we plot the situation $t'/t=0.2$ in (d).}
\label{zstart_t02}
\end{figure*}

The figure of merit is the number that is most important when it
comes to technological applications regarding
thermoelectrics, as mentioned above, being a measure of
thermoelectrical efficiency.  In our calculations we are ignoring
the lattice contribution to the Lorenz number so our
FOM calculated is that due to only electronic contributions.

Using the ``tricks'' to handle finite size effects for $S^*(T)$ and
$L^*(T)$, c.f. Ref.~\onlinecite{peterson}, we calculate
the high frequency expansion of the FOM ($Z^*(T)T$) given in Eq.~\ref{zstart}.

A large FOM can arise in essentially two ways. One way is for the thermopower,
which is squared in the numerator, to be large.  The other is for the
Lorenz number in the denominator to be small.  The Lorenz number has
been shown in Sec~\ref{sec-l} to tend to zero as the temperature
tends to infinity.  The thermopower, on the other hand, reaches a constant,
and finite, MH limit as $T\rightarrow\infty$.  Hence, the electronic contribution
to the FOM will eventually grow to infinity as the temperature is increased
without bound.

In the following subsections we plot $Z(\omega,T)T$ as a function of $T$ and
filling $n$.

\subsection{Hubbard with $t'=0$}

Fig.~\ref{zstart_t0}(a)-(c) shows the FOM for the $t'=0$ Hubbard model for the
non-interacting case, the weakly coupled case, and the strongly coupled case.  Also
 plotted is the DC limit of the full frequency dependent FOM.  In the
non-interacting case (Fig.~\ref{zstart_t0}(a)) the FOM grows
apparently quadratically in temperature with a coefficient
that decreases inversely with filling.  This behavior is seen
in the weakly coupled and strongly coupled cases as well, c.f. Fig.~\ref{zstart_t0}(b)-(c),
although the FOM is decreased more for lower fillings.

The full frequency behavior of the FOM (not shown) is very similar to the
high frequency limit, evidently because the differences in the two limits for the
thermopower and Lorenz number largely cancel one another out.
\\
\subsection{Hubbard with $t'\neq 0$}

In Fig.~\ref{zstart_t02}(a)-(c) and Fig.~\ref{zstart_t02}(d) we investigate
the effects of frustration on the FOM for the $t'<0$ and $t'>0$
cases, respectively.

For the situation that produced a low temperature enhancement
in the thermopower and Lorenz number ($t'<0$) the FOM is
suppressed compared to $t'=0$ except for the non-interacting
case where they are very similar.  The larger interaction
strength has the effect of further suppressing the FOM especially
for low fillings.

Alternatively, the case of $t'>0$ that produced suppression in
both $S^*(T)$ and $L^*(T)$ has a very slight enhancement
in the FOM shown in Fig.~\ref{zstart_t02}(d).  This nicely illustrates
the complicated way in which the thermopower and Lorenz number
combine to produce the FOM.

\section{Conclusion}\label{sec-conc}

We have computed thermoelectrical properties of the Hubbard model
and the spin-less fermion $t$-$V$ model on one-dimensional rings investigating,
in particular, the thermopower (Hubbard and $t$-$V$), Lorenz number, and figure of
merit (Hubbard only).  Our calculations are the first detailed calculations of these 
thermoelectric variables in the literature for strongly correlated models.  
By adding a second neighbor hopping term with
amplitude $t'$, both positive and negative, we were able to
destroy the integrability of these models and induce frustration.  For
$t'<0$ the Hubbard and $t$-$V$ model displayed an enhanced
thermopower at low to intermediate temperatures.  For the
Hubbard model the Lorenz number was also found to have low
temperature enhancements for $t'<0$.  However, the FOM did not
produce the same enhancement for $t'<0$
but instead a suppression for non-zero interaction strength $U$.
Although, the FOM was modestly enhanced by the opposite
hopping $t'>0$ at low fillings.

For the Hubbard model, the
thermopower had a generally weak frequency
dependence other than a sometimes large feature near $\omega \sim U$.  This
behavior was also obtained, but not shown here, for the
Lorenz number and FOM.  This has the consequence that the high frequency
versions of the thermopower, Lorenz number, and FOM recently proposed by
Shastry\cite{shastry_1,shastry_2,shastry_3} provide a good approximation
to the full dynamical quantities for most values of the system parameters.

\begin{acknowledgments}
We gratefully acknowledge support from Grant No. NSF-DMR0408247 and
DOE BES DE-FG02-06ER46319.  We acknowledge helpful conversations
with J. E. Moore.   SM thanks the DOE for support.
\end{acknowledgments}


\begin{thebibliography}{40}
\bibitem[{cl-()}]{cl-note}
\bibinfo{note}{The computational limitations for the Hubbard model are due
  mostly to the double sum over states when calculating conductivities 
  via the Kubo formulas rather than due to the diagonalization
  proper.}
\bibitem[{eps()}]{epsilon-foot}
\bibinfo{note}{The mean energy level spacing is also very insentive to the
  value of $t'$ used here.}
\bibitem[{mh-()}]{mh-limit-note}
\bibinfo{note}{The MH limit for a finite system does not exactly equal the MH
  limit for the infinite system even in the $T\rightarrow\infty$ limit due to
  the finite nature of the Hilbert space.}
\bibitem[{t02()}]{t02-note}
\bibinfo{note}{For the frustrated case, there is a frequency dependence at half
  filling, unlike the unfrustrated case, arising from the broken particle-hole
  symmetry.}
\end{thebibliography}

\end{document}